\documentclass{article}
\usepackage[dvips]{graphicx}  

\begin{document}


\begin{center}
{\Large\bf  A New  Light-Speed Anisotropy Experiment: Absolute Motion\\ and Gravitational Waves Detected\rule{0pt}{13pt}}\par
\bigskip
Reginald T. Cahill \\ 
{\small\it School of Chemistry, Physics and Earth Sciences, Flinders University,
Adelaide 5001, Australia\rule{0pt}{13pt}}\\

\raisebox{+1pt}{\footnotesize E-mail: Reg.Cahill@flinders.edu.au}\\

Published: {\it Progress in Physics},  {\bf 4}, 73-92, 2006.
\par
\bigskip\vspace*{-3pt}
{\small\parbox{11cm}{%
Data from a new experiment measuring the anisotropy of the one-way  
speed of EM waves  in a coaxial cable, gives the speed of light as  
300,000$\pm$400$\pm$20km/s in a measured direction
RA$=$5.5$\pm$2\,hrs, Dec$=$70$\pm$10$^\circ$S, is shown to be in excellent agreement with the results from seven previous anisotropy experiments, particularly those of Miller (1925/26), and even those of Michelson and Morley (1887).  The Miller gas-mode interferometer results, and those from the RF coaxial cable experiments of Torr and Kolen (1983), De Witte (1991) and the new experiment all reveal the presence of gravitational waves,  as indicated by the last $\pm$ variations above, but of a kind different from those supposedly predicted by General Relativity.  Miller repeated the Michelson-Morley 1887 gas-mode interferometer experiment and again detected the anisotropy of the speed of light, primarily in the years 1925/1926 atop Mt.Wilson, California.  The understanding of the operation of the Michelson interferometer in gas-mode was only achieved in 2002 and  involved a calibration for the interferometer that necessarily involved Special Relativity effects and the refractive index of the gas in the light paths.    The results demonstrate the reality of the Fitzgerald-Lorentz contraction as an observer independent relativistic effect.   A common misunderstanding is that the anisotropy of the speed of light is necessarily in conflict with Special Relativity and Lorentz symmetry --- this is explained.   All eight experiments and theory show that we have both anisotropy of the speed of light {\it and} relativistic effects, and that a dynamical 3-space exists ---  that absolute motion through that space has been repeatedly observed since 1887. These developments completely change  fundamental physics and our understanding of reality. ``Modern'' vacuum-mode Michelson interferometers, particularly the long baseline terrestrial versions, are, by design flaw, incapable of detecting  the anisotropy effect and  the gravitational waves.
\rule[0pt]{0pt}{0pt}}}\vspace*{-2pt}\medskip 
\end{center}

\markboth{R.\,T.\,Cahill. A New  Light-Speed Anisotropy Experiment: Absolute
Motion and Gravitational Waves Detected}{\thepage}
\markright{R.\,T.\,Cahill. A New  Light-Speed Anisotropy Experiment: Absolute
Motion and Gravitational Waves Detected}

\label{art-715}

\setcounter{equation}{0}
\setcounter{figure}{0}
\setcounter{table}{0}
\setcounter{section}{0}

\centerline{\bf Contents}

{\footnotesize

\vspace*{8pt}
\noindent
\textbf{1\rule{6pt}{0pt}Introduction}\dotfill \pageref{intro}\rule{3mm}{0pt}

\vspace*{4pt}
\noindent
\textbf{2\rule{6pt}{0pt}Special Relativity and the speed of light 
anisotropy}\dotfill \pageref{sect:SR}\rule{3mm}{0pt}

\vspace*{4pt}
\noindent
\textbf{3\rule{6pt}{0pt}Light speed anisotropy experiments}\dotfill \pageref{sect:3}\rule{3mm}{0pt}
\begin{itemize}
\vspace*{-1pt}
\item [3.1] Michelson gas-mode interferometer\dotfill \pageref{subsect:31}\rule{3mm}{0pt}

\vspace*{-2pt}
\item [3.2] Michelson-Morley experiment\dotfill \pageref{subsect:32}\rule{3mm}{0pt}
  
\vspace*{-2pt}
\item [3.3] Miller interferometer\dotfill \pageref{subsect:33}\rule{3mm}{0pt}

\vspace*{-2pt}
\item [3.4] Other gas-mode Michelson interferometer experiments\dotfill 
\pageref{subsect:34}\rule{3mm}{0pt}

\vspace*{-2pt}
\item [3.5] Coaxial cable speed of EM waves anisotropy experiments\dotfill
\pageref{subsect:35}\rule{3mm}{0pt}

\vspace*{-2pt}
\item [3.6] Torr-Kolen coaxial cable anisotropy experiment\dotfill 
\pageref{subsect:36}\rule{3mm}{0pt}
  
\vspace*{-2pt}
\item [3.7] De Witte coaxial cable anisotropy experiment\dotfill 
\pageref{subsect:37}\rule{3mm}{0pt}
\end{itemize}

\vspace*{2pt}
\noindent
\textbf{4\rule{6pt}{0pt}Flinders University gravitational wave detector}\dotfill
\pageref{sect:4}\rule{3mm}{0pt}
\begin{itemize}
\vspace*{-1pt}
\item [4.1] Optical fibre effect\dotfill \pageref{subsect:41}\rule{3mm}{0pt}

\vspace*{-2pt}
\item [4.2] Experimental components\dotfill \pageref{subsect:42}\rule{3mm}{0pt}
  
\vspace*{-2pt}
\item [4.3] All-optical detector\dotfill \pageref{subsect:43}\rule{3mm}{0pt}

\vspace*{-2pt}
\item [4.4] Results from the Flinders detector\dotfill \pageref{subsect:44}\rule{3mm}{0pt}

\vspace*{-2pt}
\item [4.5] Right ascension\dotfill
\pageref{subsect:45}\rule{3mm}{0pt}

\vspace*{-2pt}
\item [4.6] Declination and speed\dotfill \pageref{subsect:46}\rule{3mm}{0pt}
  
\vspace*{-2pt}
\item [4.7] Gravity and gravitational waves\dotfill \pageref{sect:gravity}\rule{3mm}{0pt}
\end{itemize}

\vspace*{2pt}
\noindent
\textbf{5\rule{6pt}{0pt}Conclusions}\dotfill \pageref{conclusions}\rule{3mm}{0pt}
\centerline{\rule{0pt}{10pt}\rule{72pt}{0.4pt}}

}

\section{Introduction}\label{intro}

\markright{R.\,T.\,Cahill. A New  Light-Speed Anisotropy Experiment: Absolute
Motion and Gravitational Waves Detected}

\vspace*{-1pt}
Of fundamental importance to physics is whether the speed of light is the same in all directions, as measured say in a  laboratory attached to the Earth. This is what is meant by {\it light speed anisotropy} in the title of this paper. The pre\-vail\-ing belief system in physics has it that the speed of light is isotropic, that there is no preferred frame of reference, that absolute motion has never been observed, and that 3-space does not, and indeed cannot exist. This is the essence of Ein\-stein's 1905 postulate that the speed of light is independent of the choice of observer.  This postulate has determined the course of physics over the last 100 years. 

Despite the enormous  significance  of this postulate there has never been a reliable direct  experimental test, that is, in which the one-way travel time of light in vacuum over a set distance has been measured, and repeated for different directions.  So how could a science as fundamental and important as physics permit such a key idea to go untested? And what are the consequences for fundamental physics if indeed, as reported herein and elsewhere, that the speed of light is anisotropic, that a dynamical 3-space does exist? This would imply that if reality is essentially space and matter, with time tracking process and change, then physics has completely missed the existence of that space.  If this is the case then this would have to be the biggest  blunder ever in the history of science, more so because some  physicists have independently  de\-tect\-ed that anisotropy. While herein we both summarise seven previous detections of the anisotropy and report a new ex\-per\-i\-ment, the implications for fundamental physics have already been substantially worked out. It leads to a new modelling and comprehension of reality known as {\it Process Physics} \cite{C11}.

The failure of mainstream physics to understand that  the speed of light is anisotropic, that a dynamical 3-space exists, is caused by an ongoing failure to comprehend the operation of the Michelson interferometer, and also by theoretical phys\-icists not understanding that the undisputed successes of special relativity effects, and even Lorentz symmetry, do not imply that the speed of light must be isotropic --- this is a mere abuse of logic, as explained later.

The Michelson interferometer is actually a complex in\-st\-rument. The problem is that the anisotropy of the speed of light affects its actual dimensions and hence its operation:  there are actual length contractions of its physical arms. Because the anisotropy of the speed of light is so fund\-a\-mental it is actually very subtle to design an effective exper\-i\-ment because the sought for effect also affects the instrument in more than one way. This subtlety has been overlooked for some 100 years, until in 2002 the original data was reanalysed using a relativistic theory for the calibration  of  the interferometer \cite{C2}.

The new understanding of the operation of the Michelson interferometer  is that it can only detect the light speed an\-iso\-tropy when there is gas in the light paths, as there was in the early experiments. Modern versions have removed the gas and made the instrument totally unable to detect the light speed anisotropy. Even in gas mode the interferometer is a very insensitive device, being 2nd order in $v/c$ and further suppressed in sensitivity by the gas refractive index dependency.

More direct than the Michelson interferometer, but still not a direct measurement, is to measure the one-speed of radio frequency (RF) electromagnetic waves in a coaxial cable, for this permits electronic timing methods. This ap\-proach is 1st order in $v/c$, and independent  of the refractive index suppression effect.  Nevertheless because it is one-way clocks are required at both ends, as in the Torr and Kolen, and De Witte  experiments, and the required length of the coaxial cable was determined, until now, by the stability of atomic clocks over long durations.

The new one-way RF coaxial experiment  reported herein utilises a new timing technique that avoids the need for  two atomic clocks, by using a very special property of optical fib\-res, namely that the light speed in optical fibres is isotropic, and is used for transmitting timing information, while in the coaxial cables the RF speed is  anisotropic, and is used as the sensor. There is as yet no explanation for this optical fibre effect, but it radically changes the technology for anisotropy experiments, as well and at the same time that of gravitation\-al wave detectors.  In  the near future all-optical gravitational wave detectors are possible in desk-top instruments.  These gravitational waves have very different properties from those supposedly predicted from General  Relativity, although that appears to be caused by errors in that derivation. 

As for gravitational waves, it has been realised now that they were seen in the Miller, Torr and Kolen, and De Witte experiments, as they are again observed in the new experiment.  Most amazing is that these wave effects also ap\-pear to be present in the Michelson-Morley fringe shift data from 1887, as the fringe shifts varied from day to day.   So Michelson and Morley should have reported that they had discovered absolute motion, a preferred frame, and also wave effects of that frame, that the speed of light has an anisotropy that fluctuated over and above that caused by the rotation of the Earth.

The first and very successful attempt to look for a preferred frame
was by Michelson and Morley in 1887. They did in fact detect the expected anisotropy at the level of $\pm$8km/s \cite{C1},  but only according to Michelson's calibration theory.   However this result has essentially been ignored ever since as they expected to detect an effect of at least $\pm$30km/s, which is the orbital speed of the earth about the sun.  As Miller recognised the basic problem with the Michelson interferometer is that the calibration of the instrument was then clearly not correctly understood, and most likely wrong \cite{C4}. Basically Michelson had used Newtonian physics to calibrate his instrument, and of course we now know that that is completely inappropriate as relativistic effects play a critical role as the interferometer is a 2nd order device ($\sim v^2/c^2$ where $v$ is the speed of the device relative to a physical dynamical 3-space\footnote{In Michelson's era the idea was that $v$ was the speed of light relative to an {\it ether}, which itself filled space.  This dualism has proven to be wrong.}), and so various effects at that order must be taken into account in determining the calibration of the instrument, that is, what light speed anisotropy corresponds to the observed fringe shifts. It was only in 2002 that the calibration of the Michelson interferometer was finally determined by taking account of  relativistic effects \cite{C2}. One aspect of that was the discovery that only a Michelson interferometer in gas-mode could detect the light anisotropy, as discussed below.  As well the interferometer when used in air is nearly a factor of 2000 less sensitive than that according to the inappropriate Newtonian theory. This meant that the Michelson and Morley anisotropy speed variation  was now around 330km/s on average, and as high as 400km/s  on some days.  Miller was aware of this calibration problem, and resorted to a brilliant indirect method, namely to observe the fringe shifts over a period of a year, and to use the effect of the earth's orbital speed upon the fringe shifts to arrive at a calibration. The earth's orbital motion was clearly evident in Miller's data, and using this effect he obtained a light speed anisotropy effect of some 200km/s in a particular direction.  However even this method made assumptions which are now known to be invalid, and correcting his earth-effect calibration method we find that it agrees with  the new relativistic effects calibration, and both methods now give a speed of near 400km/s. This also then agrees with the Michelson-Morley results. Major discoveries like that of Miller must be reproduced by different experiments and by different techniques.  Most significantly there are in total  seven other experiments that confirm this Miller result, with four being gas-mode Michelson interferometers using either air, helium or a He/Ne mixture in the light path, and three experiments that measure variations in the   one-way speed of EM waves travelling through a coaxial cable as the orientation of the cable is changed, with the latest being a high precision technique reported herein and in \cite{Salz,LA}.  This method is 1st order in $v/c$, so it does not  require  relativistic effects to be taken into account, as discussed later.

As the Michelson interferometer requires a gas to be present in the light path in order to detect the anisotropy it follows that vacuum interferometers, such as those in \cite{C3},  are simply inappropriate for the task, and   it is surprising that some attempts to detect the anisotropy  in the speed of light still use vacuum-mode Michelson interferometers, some years after the 2002 discovery of the need for a gas in the light path \cite{C2}.  

Despite the extensive data collected and analysed by Miller after his fastidious testing and refinements to control temperature effects and the like, and most importantly his demonstration that the effects tracked sidereal time and not solar time,  the world of physics has, since publication of the results by MIller  in 1933, simply ignored this discovery. The most plausible explanation for this  situation is the ongoing misunderstanding by many physicists, but certainly not all,  
that any anisotropy in the speed of light must necessarily by  
incompatible with Special Relativity (SR), with SR certainly well  
confirmed experimentally.  This is misunder\-stand\-ing is clarified.  In fact Miller's data can now be used to confirm an important aspect of SR. Even so, ignoring the results of a major experiment simply because they challenge a prevailing belief system is not science --- ignoring the Miller experiment has stalled physics for some 70 years.

It is clear that the Miller experiment was highly success\-ful and highly significant, and we now know this because the same results have been obtained by later experiments which used {\it different}  experimental techniques.    The most significant part of Miller's rigorous experiment was that he showed that the effect tracked sidereal time and not solar time --- this is the acid test which shows that the direction of the anisotropy velocity vector is relative to the stars and not to the position of the Sun.  This difference is only some 4 minutes per day, but over a year amounts to a huge 24 hours  effect, and Miller saw that effect and extensively discussed it in his paper.  Similarly  De Witte in his extensive 1991 coaxial cable experiment \cite{DeWitte} also took data for 178 days to again establish the sidereal time effect: over 178 days this effect amounts to a shift in the phase of the signal through some 12 hours!   The sidereal effect  has also been established in the new coaxial cable experiment  by the author from data spanning some 200 days.

The interpretation that has emerged from the Miller and re\-lat\-ed discoveries is that space exists, that it is an  observ\-able and dynamical system, and that the Special Relativity effects \rule{-.3pt}{0pt}are \rule{-.3pt}{0pt}caused \rule{-.3pt}{0pt}by \rule{-.3pt}{0pt}the \rule{-.3pt}{0pt}absolute \rule{-.3pt}{0pt}motion \rule{-.3pt}{0pt}of \rule{-.3pt}{0pt}quantum \rule{-.3pt}{0pt}systems through that space \cite{C11,BHoles}.  This is essentially the Lo\-rentz~in\-ter\-pretation of Special Relativity, and then the space\-time is me\-re\-ly a mathematical construct.  The new understanding has lead to an explanation of why Lorentz symmetry  manifests despite there being a preferred frame, that is,  a local frame in which only therein is the speed of light  isotropic.  A minimal theory for the dynamics of this space has been developed \cite{C11,BHoles}  which has resulted in an explanation of numerous~phe\-nom\-ena, such as gravity as a quantum effect \cite{BHoles,Schrod}, the~so-called ``dark matter'' effect, the black hole systematics, gravi\-tational light bending, gravitational lensing, and so~[21--25].

    The Miller data also revealed another major discovery that Miller himself may not have understood, namely that the anisotropy vector actually fluctuates form hour to hour and day to day even when we remove the manifest effect of the Earth's rotation, for Miller may have interpreted this as being caused by imperfections in his experiment. This means that the flow of space past the Earth displays turbulence or a wave effect: basically the Miller data has revealed what we now call {\it gravitational waves},  although these are different to the waves supposedly predicted by General Relativity.  These wave effects were also present in the  Torr and Kolen \cite{C8}   first coaxial cable experiment at Utah University in 1981, and were again manifest in the  De Witte data  from 1991. Analysis of the De Witte data has shown that these waves have a fractal structure \cite{DeWitte}.  The Flinders University Grav\-i\-tational Waves Detector (also a coaxial cable experiment) was constructed to investigate these waves effects. This  sees the wave effects detected by Miller, Torr and Kolen, and by De Witte.   The plan of this paper is to first outline the modern understanding of how a gas-mode Michelson interferometer actually operates, and the nature, accuracy and significance of the Miller experiment. We also report the other seven experiments that confirm the Miller discoveries, particularly data from the new high-precision gravity wave detector that detects not only a light speed anisotropy but also the wave effects.

\markright{R.\,T.\,Cahill. A New  Light-Speed Anisotropy Experiment: Absolute
Motion and Gravitational Waves Detected}

\section{Special Relativity and the speed of light anisotropy\label{sect:SR}}

\markright{R.\,T.\,Cahill. A New  Light-Speed Anisotropy Experiment: Absolute
Motion and Gravitational Waves Detected}

\vspace*{-1pt}
It is often assumed that the anisotropy of the speed of light  is inconsistent with Special Relativity, that only one or the 
other can be valid, that they are mutually incompatible.  This misunderstanding is very prevalent in the literature of phys\-ics, although   this conceptual error has been explained \cite{C11}.  The error is based upon a misunderstanding of how the logic of theoretical physics works, namely the important difference between an {\it if} statement, and an  {\it if and only if} statement.   To see how this confusion has arisen we need to recall the history of Special Relativity (SR). In 1905 Einstein deduced the SR formalism by assuming, in part, that the speed of light is invariant  for all relatively moving observers, although most importantly one must ask just how that speed is defined or is to be measured. The SR formalism then predicted numerous effects,  which have been ex\-ten\-sively confirmed by experiments over the last 100 years.    However this Einstein derivation was an 
{\it if}  statement, and not an {\it if and only if}  statement.  For an {\it if} statement, that {\it if A then B}, does not imply the truth of {\it A} if {\it B} is found to be  true; only an {\it if and only if }  statement has that property, and Einstein did not construct such an argument.
What this means is that the validity of the various SR effects does {\it not} imply that the speed of light must be isotropic.  This is actually implicit in the SR formalism itself, for it permits one to use any particular foliation of the 4-dimensional spacetime into a 3-space and a 1-space (for time). Most importantly it does not forbid  that one particular foliation be actual.  So to analyse the data from gas-mode interferometer experiments we must use the SR effects, and the fringe shifts   reveal  the preferred frame, an actual 3-space, by revealing the anisotropic speed of light, as Maxwell and Michelson had originally believed. 

For ``modern'' resonant-cavity Michelson interferometer experiments we predict no rotation-induced fringe shifts, unless operated in gas-mode.  Unfortunately in analysing the data from the vacuum-mode experiments the consequent null effect is misinterpreted, as in \cite{C3}, to imply the absence of a preferred direction, of absolute motion. But it is absolute motion which
 causes the dynamical effects of length contrac\-tions, time dilations and other relativistic effects, in accord with  Lorentzian interpretation of relativistic effects.  
 
 The detection of absolute motion is not incompatible with  Lorentz symmetry; the contrary belief was postulated by Einstein, and has persisted for over 100 years, since 1905. So far the experimental evidence is that absolute motion and Lorentz symmetry are real and valid phenomena; absolute motion is motion presumably relative to some substructure to space, whereas Lorentz symmetry parameterises dynamical effects caused
by the motion of systems through that sub\-struc\-ture. To check Lorentz symmetry we can use vacuum-mode resonant-cavity interferometers, but using gas within the resonant-cavities would enable 
these devices to detect absolute motion with great precision. As well there are novel wave phenomena that could also  be studied, as discussed herein and in  \cite{C9,C10}.

Motion through the structured space, it is argued, induces actual dynamical time dilations and length
contractions in agree\-ment with the Lorentz interpretation of special relati\-vist\-ic
effects.  Then observers in  uniform motion ``through'' the space will, on measurement  of the speed of light  using the special but misleading Einstein measurement protocol, obtain
always the same numerical value  $c$.   To see this expli\-citly consider how various observers $P, P^\prime,\dots$
moving with  different  speeds through space, measure the speed of light.  They  each acquire a standard rod 
and an accompanying stan\-d\-ardised clock. That means that these standard  rods  would agree if they were brought
together, and at rest with respect to space they would all have length $\Delta l_0$, and similarly for
the clocks.    Observer $P$ and accompanying rod are both moving at  speed $v_R$ relative to space, with
the rod longitudinal to that motion. P  then  measures the time
$\Delta t_R$, with the clock at end $A$ of the rod,  for a light pulse to travel from  end $A$ to the other end
$B$  and back again to $A$. The  light  travels at speed $c$ relative to space. Let the time taken for
the light pulse to travel from
$A\,{\rightarrow}\, B\,$ be $t_{AB}$ and  from $B\,{\rightarrow}\, A$ be $t_{BA}$, as measured by a clock at rest with respect
to space\footnote{Not all clocks will behave in this same ``ideal'' manner.}$\!$. The  length of the rod 
moving at speed
$v_R$ is contracted to 
\vspace*{-3pt}
\begin{equation}
\Delta l_R=\Delta l_0\,\sqrt{1-\frac{v_R^2}{c^2}}\,.
\label{eqnum:c0}\end{equation}

\vspace*{-2pt}
In moving from  $A$ to $B$ the light must travel an extra  distance 
because the  end  $B$ travels a distance $v_Rt_{AB}$ in this time, thus the total distance that must be
traversed  is
\vspace*{-2pt}
\begin{equation}\label{eqnum:c1}
ct_{AB}=\Delta l_R+v_R\,t_{AB}\,,
\end{equation}

\vspace*{-2pt}\noindent
similarly on returning from $B$ to $A$ the light must travel the distance
\vspace*{-2pt}
\begin{equation}\label{eqnum:c2}
ct_{BA}=\Delta l_R-v_R\,t_{BA}\,.
\end{equation}

\vspace*{1pt}
Hence the total travel time $\Delta t_0$ is
\begin{eqnarray}\label{eqnum:c3}
\Delta t_0=t_{AB}+t_{BA}&=&\frac{\Delta l_R}{c-v_R}+\frac{\Delta l_R}{c+v_R}=\\[+4pt]
&=&\frac{2\Delta l_0}{c\,\sqrt{1-\displaystyle\frac{v_R^2}{c^2}}}\,.
\end{eqnarray}

\vspace*{-5pt}
Because  of  the time dilation effect for the moving clock
\vspace*{-3pt}
\begin{equation}
\Delta t_R=\Delta t_0\,\sqrt{1-\displaystyle\frac{v_R^2}{c^2}}\,.
\label{eqnum:c4}\end{equation}

\vspace*{-2pt}
Then for the moving observer the speed of light is de\-fin\-ed as the distance the observer believes the light
travelled ($2\Delta l_0$) divided by the travel time according to the accom\-pa\-nying clock ($\Delta t_R$), namely 
$2\Delta l_0/\Delta t_R = 2\Delta l_R/\Delta t_0$, from above, which is thus the same speed as seen by an observer at rest in the space, namely $c$.  So the speed $v_R$ of the ob\-server through space is not revealed by this
pro\-cedure, and the observer is erroneously led to the conclusion that the speed of light is always $c$. 
This follows from two or more observers in manifest relative motion all obtaining the same speed c by this
procedure. Despite this failure  this special effect is actually the basis of the spacetime\index{spacetime}
Einstein mea\-su\-re\-ment protocol. That this protocol is blind to the ab\-sol\-ute motion has led to enormous confusion within physics.

To be explicit the Einstein measurement protocol\index{measurement protocol} actual\-ly inadvertently uses this
special effect by using the radar method for assigning historical spacetime coordinates to an event: the observer
records the time of emission and recep\-tion of radar pulses ($t_r \,{>}\, t_e$) travelling through  space, and then retrospectively assigns the time and distance of a distant event
$B$ according to (ignoring directional information for simplicity) 
\vspace*{-6pt}
\begin{equation}T_B=\frac{1}{2}\,\bigl(t_r+t_e\bigr)\,, \qquad
D_B=\frac{c}{2}\,\bigl(t_r-t_e\bigr)\,,\label{eqnum:25}\end{equation}
%
where each observer is now using the same numerical value of $c$.
 The event $B$ is then plotted as a point in 
an individual  geometrical construct by each  observer,  known as a space\-time record, with coordinates $(D_B,T_B)$. This
is no different to an historian recording events according to  some agreed protocol.  Unlike historians, who
don't confuse history books with reality, physicists do so. 
  We now show that because of this
protocol and the absolute motion dynamical effects, observers will discover on comparing their
historical records of the same events that the expression
\vspace*{-1pt}
\begin{equation}
 \tau_{AB}^2 =   T_{AB}^2- \frac{1}{c^2} \,D_{AB}^2\,,
\label{eqnum:26}\end{equation}

\vspace*{-1pt}\noindent
is an invariant, where $T_{AB}=T_A-T_B$ and $D_{AB}=D_A-D_B$ are the differences in times and distances
assigned to events $A$ and
$B$ using the Einstein measurement protocol (\ref{eqnum:25}), so long as both are sufficiently small
compared with the scale of inhomogeneities  in the velocity field.

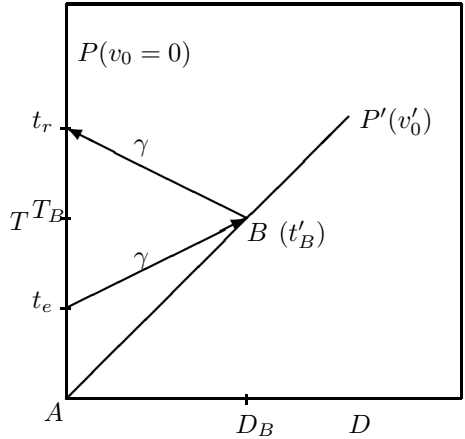
\begin{figure}[t]
\vspace{-14.2mm}
\hspace{+20mm}
\setlength{\unitlength}{1.5mm}
\hspace{30mm}\begin{picture}(40,45)
\thicklines
\put(-2,-2){{\bf $A$}}
\put(+1,30){{\bf $P(v_0=0)$}}
\put(16,14){\bf $B$ $(t^\prime_B)$}
\put(25,-3){\bf $D$}
\put(15,-3){\bf $D_B$}
\put(16,-0.5){\line(0,1){0.9}}
\put(-5,15){\bf $T$}
\put(26,24){$ P^\prime(v^\prime_0$)}

\put(0,0){\line(1,0){35}}
\put(0,0){\line(0,1){35}}
\put(0,35){\line(1,0){35}}
\put(35,0){\line(0,1){35}}
\put(0,0){\line(1,1){25}}

\put(0,8){\vector(2,1){16}}
\put(16,16){\vector(-2,1){16}}

\put(-3,8){\bf $t_e$}\put(-0.5,8){\line(1,0){0.9}}
\put(-3,16){\bf $T_B$}\put(-0.5,16){\line(1,0){0.9}}
\put(-3,24){\bf $t_r$}\put(-0.5,24){\line(1,0){0.9}}
\put(6,12){$\gamma$}
\put(6,22){$\gamma$}

\end{picture}
\vspace*{2mm}
\caption{\small  Here $T\,{-}\,D$ is the spacetime construct (from  the Einstein measurement protocol) of a special observer
$P$ {\it at rest} wrt space, so that $v_0\,{=}\,0$.  Observer $P^\prime$ is moving with speed
$v^\prime_0$ as determined by observer $P$, and therefore with speed $v^\prime_R=v^\prime_0$ wrt space. Two light
pulses are shown, each travelling at speed $c$ wrt both $P$ and space.   Event
$A$ is when the observers pass, and is also used to define zero time  for each for
convenience. }\label{fig:spacetime1}

\vspace*{-2mm}
\end{figure}

To confirm the invariant  nature of the construct in   (\ref{eqnum:26}) one must pay careful attention to
observational times as distinct from protocol times and distances, and this must be done separately for each
observer.  This can be tedious.  We now  demonstrate this for the situation illustrated in
Fig.~\ref{fig:spacetime1}. 

 By definition  the speed of
$P^\prime$ according to
$P$ is
$v_0^\prime =D_B/T_B$ and so
$v_R^\prime=v^\prime_0$,  where 
$T_B$ and $D_B$ are the protocol time and distance for event $B$ for observer $P$ according to
(\ref{eqnum:25}).  Then using (\ref{eqnum:26})  $P$ would find that
$(\tau^P_{AB})^2=T_{B}^2-\frac{1}{c^2}D_B^2$ since both
$T_A=0$ and $D_A$=0, and whence $(\tau^{P}_{AB})^2=(1-\frac{v_R^{\prime 2}}{c^2})T_B^2=(t^\prime_B)^2$ where
the last equality follows from the time dilation effect on the $P^\prime$ clock, since $t^\prime_B$ is the time
of event
$B$ according to that clock. Then $T_B$ is also the time that $P^\prime$  would compute for event $B$ when
correcting for the time-dilation effect, as the speed $v^\prime_R$ of $P^\prime$ through the quantum foam is
observable by $P^\prime$.  Then $T_B$ is the `common time' for event $B$ assigned by both
observers.  For
$P^\prime$ we obtain  directly, also from  (\ref{eqnum:25}) and (\ref{eqnum:26}), that
$(\tau^{P'}_{AB})^2=(T_B^\prime)^2-\frac{1}{c^2}(D^\prime_B)^2=(t^\prime_B)^2$, as $D^\prime_B=0$  and
$T_B^\prime=t^\prime_B$. Whence for this situation
\begin{equation}
(\tau^{P}_{AB})^2=(\tau^{P'}_{AB})^2,
\label{eqn:invariant1}
\end{equation} and so the
 construction  (\ref{eqnum:26})  is an invariant.

While so far we have only established the invariance of the construct  (\ref{eqnum:26}) when one of the
observers is at rest in space, it follows that for two observers $P^\prime$ and
$P^{\prime\prime}$ both in absolute motion  it follows that they also agree on the invariance
of (\ref{eqnum:26}).  This is easily seen by using the intermediate step of  a stationary observer $P$:
\begin{equation}
(\tau^{P'}_{AB})^2=(\tau^{P}_{AB})^2=(\tau^{P''}_{AB})^2.
\label{eqn:invariant2}
\end{equation}

\begin{figure*}[t]
\setlength{\unitlength}{0.9mm}
\hspace{34mm}\begin{picture}(0,30)
\thicklines
\put(-10,0){\line(1,0){50}}
\put(-5,0){\vector(1,0){5}}
\put(40,-1){\line(-1,0){29.2}}
\put(15,0){\vector(1,0){5}}
\put(30,-1){\vector(-1,0){5}}
\put(10,0){\line(0,1){30}}
\put(10,5){\vector(0,1){5}}
\put(11,25){\vector(0,-1){5}}
\put(11,30){\line(0,-1){38}}
\put(11,-2){\vector(0,-1){5}}
\put(8.0,-2){\line(1,1){5}}
\put(9.0,-2.9){\line(1,1){5}}
\put(6.5,30){\line(1,0){8}}
\put(40,-4.5){\line(0,1){8}}
\put(5,12){ $L$}
\put(4,-5){ $A$}
\put(35,-5){ $B$}
\put(25,-5){ $L$}
\put(12,26){ $C$}
\put(9,-8){\line(1,0){5}}
\put(9,-9){\line(1,0){5}}
\put(14,-9){\line(0,1){1}}
\put(9,-9){\line(0,1){1}}
\put(15,-9){ $D$}
\put(50,0){\line(1,0){50}}
\put(55,0){\vector(1,0){5}}
\put(73,0){\vector(1,0){5}}
\put(85,0){\vector(1,0){5}}
\put(90,15){\vector(1,0){5}}
\put(100,-1){\vector(-1,0){5}}
\put(100,-4.5){\line(0,1){8}}
\put(68.5,-1.5){\line(1,1){4}}
\put(69.3,-2.0){\line(1,1){4}}
\put(70,0){\line(1,4){7.5}}
\put(70,0){\vector(1,4){3.5}}
\put(77.5,30){\line(1,-4){9.63}}
\put(77.5,30){\vector(1,-4){5}}
\put(73.5,30){\line(1,0){8}}
\put(83.3,-1.5){\line(1,1){4}}
\put(84.0,-2.0){\line(1,1){4}}
\put(100,-1){\line(-1,0){14.9}}
\put(73,3){$\alpha$}
\put(67,-5){ $A_1$}
\put(80,-5){ $A_2$}
\put(85,-8){\line(1,0){5}}
\put(85,-9){\line(1,0){5}}
\put(90,-9){\line(0,1){1}}
\put(85,-9){\line(0,1){1}}
\put(90,-9){ $D$}
\put(95,-5){ $B$}
\put(79,26){ $C$}
\put(90,16){ $v$}
\put(-8,8){(a)}
\put(55,8){(b)}

\end{picture}

\vspace{7mm}
\caption{\small{Schematic diagrams of the Michelson Interferometer, with beamsplitter/mirror at $A$ and
mirrors at $B$ and $C$ on arms  from $A$, with the arms of equal length $L$ when at rest.  $D$ is a 
screen or detector. In (a) the interferometer is
at rest in space. In (b) the interferometer is moving with speed $v$ relative to space in the direction
indicated. Interference fringes are observed at the  detector $D$.  If the interferometer is
rotated in the plane  through $90^o$, the roles of arms $AC$ and $AB$ are interchanged, and during the
rotation shifts of the fringes are seen in the case of absolute motion, but only if the apparatus operates
in a gas.  By counting fringe changes the speed $v$ may be determined.}  \label{fig:Minterferometer}}

\vspace*{2.8mm}
\end{figure*}
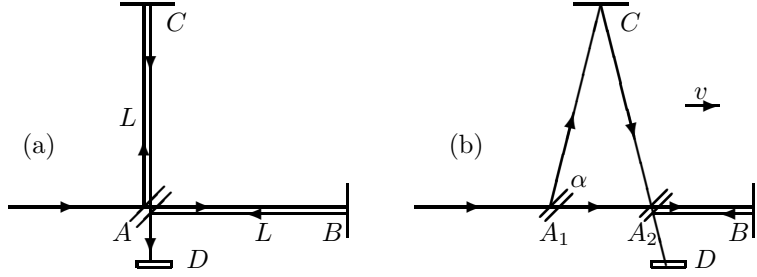

Hence the protocol and Lorentzian absolute motion ef\-fects result in the construction in (\ref{eqnum:26})  being indeed an
invar\-iant\index{invariant interval} in general.  This  is  a remarkable and subtle result.  For Einstein this
invariance was a fundamental assumption, but here it is a derived result, but one which is nevertheless deeply
misleading. Explicitly indicating  small quantities  by $\Delta$ prefixes, and on comparing records
retrospectively, an ensemble of nearby observers  agree on the invariant
\vspace*{-2pt}
\begin{equation}
\Delta \tau^2=\Delta T^2-\frac{1}{c^2}\,\Delta D^2,
\label{eqnum:31}\end{equation} 

\vspace*{-2pt}\noindent
for any two nearby events.  This implies that their individual patches of spacetime records may be mapped one
into the other merely by a change of coordinates, and that collecti\-ve\-ly the spacetime patches  of all may
be represented by one pseudo-Riemannian manifold, where the choice of coord\-i\-na\-tes for this manifold is
arbitrary, and we finally arrive at the invariant 
\vspace*{-2pt}
\begin{equation}
\Delta\tau^2=g_{\mu\nu}(x)\,\Delta x^\mu \Delta x^\nu,
\label{eqnum:inv}\end{equation} 

\vspace*{3pt}\noindent
with $x^\mu=\{D_1,D_2,D_3,T\}$.  Eqn.~(\ref{eqnum:inv})  is invariant under the Lorentz transformations
\begin{equation}
x^{\prime\mu}={L^\mu}_{\!\nu}\, x^\nu,\rule[-8pt]{0pt}{0pt}
\label{eqnum:Lorentz}\end{equation}
where, for example for relative motion in the $x$ direction, ${L^\mu}_{\!\nu}$ is specified by
\vspace*{-4pt}
\begin{equation}
\begin{array}{ll}
\displaystyle
x^\prime=\frac{x-vt}{\sqrt{1-v^2/c^2}}\,, \\
\displaystyle
y^\prime=y\,,\rule{0pt}{13pt}  \\
\displaystyle
z^\prime=z\,,\rule{0pt}{17pt}  \\
\displaystyle
t^\prime=\frac{t-vx/c^2}{\sqrt{1-v^2/c^2}}\,.\rule{0pt}{19pt}
\label{eqnum:xLorentz}\end{array}
\end{equation}

So absolute motion and special relativity effects, and even Lorentz symmetry, are all compatible:  a possible pre\-ferr\-ed frame is  hidden by the Einstein measurement protocol. 

So the experimental question is then whether or not a supposed preferred frame actually exists  or not --- can it be detected experimentally?  The answer is that there are now eight such consistent experiments.  In Sect.~\ref{sect:gravity}  we generalise the Dirac equation to take account of the coupling of the spinor to an actual dynamical space. This reveals again that relativistic effects are consistent with a preferred frame --- an actual space. Furthermore this leads to the first derivation of gravity from a deeper theory --- gravity turns out to be a quantum matter wave effect. 

\markright{R.\,T.\,Cahill. A New  Light-Speed Anisotropy Experiment: Absolute
Motion and Gravitational Waves Detected}

\section{Light speed anisotropy experiments}\label{sect:3}

\markright{R.\,T.\,Cahill. A New  Light-Speed Anisotropy Experiment: Absolute
Motion and Gravitational Waves Detected}

We now  consider the various experiments from over more than 100 years that have detected the anisotropy of the speed of light, and so the existence of an actual dynamical space, an observable preferred frame.  As well the experiments, it is now understood, showed that this frame is dynamical, it exhibits time-dependent effects, and that these are ``grav\-i\-ta\-tional waves''.

\vspace*{-2pt}
\subsection{Michelson gas-mode interferometer}\label{subsect:31}

Let us first consider the new understanding of how the Mich\-elson interferometer works.  This brilliant but very subtle device was conceived by Michelson as  a means to detect the anisotropy of the speed of light, as was expected towards the end of the 19th century.   Michelson used Newtonian physics to develop the theory and hence the calibration \rule{-.2pt}{0pt}for \rule{-.2pt}{0pt}his \rule{-.2pt}{0pt}device.  However we now understand that this  device  detects 2nd order effects in $v/c$ to determine $v$, and  so we must use re\-la\-tivistic effects.   However the application and analysis of data from various Michelson interferometer experiments  using a relativistic theory only occurred in 2002, some  97 years after the development of Special Relativity by Einstein, and some 115 years after the famous 1887 experiment.  As a consequence of the necessity of  using relativistic effects it was discovered in 2002 that the gas in the light paths plays a critical role, and that we finally understand how to calibrate the device, and we also discovered,  some  76 years after the 1925/26 Miller experiment, what determines the calibration constant \rule{-.2pt}{0pt}that \rule{-.2pt}{0pt}Miller \rule{-.2pt}{0pt}had \rule{-.2pt}{0pt}determined \rule{-.2pt}{0pt}using \rule{-.2pt}{0pt}the \rule{-.2pt}{0pt}Earth's \rule{-.2pt}{0pt}rotation speed about the Sun to set the calibration.   This, as we discuss later, has enabled us to now appreciate that gas-mode Mich\-elson interferometer experiments have confirmed the reality of the Fitzgerald-Lorentz length contraction effect:  in the usual interpretation of Special Relativity this effect, and others, is usually regarded as an observer dependent effect, an illusion induced by the spacetime. But the experiments are to the contrary showing that the length contraction effect is an actual observer-independent dynamical effect, as Fitz\-gerald \cite{Fitzgerald} and Lorentz had proposed \cite{Lorentz}.

The Michelson interferometer compares the change in the difference between
travel times, when the device is ro\-tat\-ed, for two coherent beams of light that travel in
orthogonal directions between mirrors; the  changing time difference being indicated by the
shift of the interference fringes during the rotation.  This effect is caused by the
absolute motion of the device through 3-space with speed $v$, and that the speed of light is
relative to that 3-space, and not relative to the apparatus/observer. However to detect the
speed of the apparatus through that 3-space gas must be present in the light paths for
purely technical reasons. The post relativistic-effects theory for this device is remarkably
simple.  The re\-la\-tivistic Fitzgerald-Lorentz contraction effect causes the arm $AB$ parallel to the
absolute velocity to be physically con\-tract\-ed to length
\vspace*{-10pt}
\begin{equation}
L_{||}=L\,\sqrt{1-\frac{v^2}{c^2}}\,.
\label{eqnum:e1}\end{equation}

The time $t_{AB}$ to travel $AB$ is set by $Vt_{AB}=L_{||}+vt_{AB}$, while for $BA$ by 
$Vt_{BA}=L_{||}-vt_{BA}$, where $V=c/n$   is the speed of light,  with
$n$  the refractive index of the gas present (we ignore here the Fresnel
drag effect for simplicity, an effect caused by the gas also being in absolute motion, see \cite{C11}).
For the total $ABA$ travel time we then obtain
\vspace*{-1pt}
\begin{equation}
t_{ABA}=t_{AB}+t_{BA}=\frac{2 LV}{V^2-v^2}\sqrt{1-\frac{v^2}{c^2}}\,.
\label{eqnum:e2}\end{equation} 

\vspace*{-1pt}
For travel in the $AC$ direction we have, from the  Pyth\-a\-goras theorem for the
 right-angled triangle in Fig.~1 that $(Vt_{AC})^2=L^2+(vt_{AC})^2$  and that
$t_{CA}=t_{AC}$. Then for the total $ACA$ travel time
\vspace*{-2pt}
\begin{equation}
t_{ACA}=t_{AC}+t_{CA}=\frac{2L}{\sqrt{V^2-v^2}}\,.
\label{eqnum:e3}\end{equation} 

\vspace*{-3pt}
 Then the difference in travel time is
\begin{equation}
\Delta t=\frac{(n^2-1)\,L}{c}\frac{v^2}{c^2}+\mbox{O}\left(\frac{v^4}{c^4} \right).
\label{eqnum:e4}\end{equation}     

\vspace*{-2pt}\noindent
after expanding in powers of $v/c$. This clearly shows that the interferometer
can only operate as a detector of absolute mo\-tion when not in vacuum ($n\,{=}\,1$), namely when
the light pass\-es through a gas, as in the early experiments (in transpa\-r\-ent solids a more
complex phenomenon occurs). A more general analysis \cite{C11}, including Fresnel drag, gives
\begin{equation}
\Delta t=k^2\frac{Lv_P^2}{c^3}\cos
\bigl(2 (\theta-\psi)\bigr) ,
\label{eqnum:e5}\end{equation}
where $k^2\,{\approx}\, n(n^2\,{-}\,1)$, while neglect of the relativistic Fitz\-gerald-Lorentz contraction effect
gives   $k^2\,{\approx}\, n^3\,{\approx }\,1$
 for 
 gases, which is essentially the Newtonian theory that Mich\-elson used.
 
 However the above analysis does not correspond to how the interferometer is actually operated. That analysis does not actually predict fringe shifts for the  field of view would be uniformly illuminated, and the observed effect would be a changing level of luminosity rather than fringe shifts. As Mi\-l\-ler knew the mirrors must be made slightly non-orthogonal, with the degree of non-orthogonality determining how many fringe shifts were visible in the field of view. Miller exper\-i\-mented with this effect to determine a comfortable number of fringes: not too few and not too many.  Hicks \cite{Hicks} deve\-lop\-ed a theory for this effect --- however it is not necessary to be aware of this analysis in using the interferometer: the non-orthogonality   reduces the symmetry of the device, and instead of having period of 180$^\circ$ the symmetry now has a period of 360$^\circ$, so that to (\ref{eqnum:e5})  we must add the extra term in
\vspace*{-4pt}
 \begin{equation}
\Delta t=k^2\frac{Lv_P^2}{c^3}\cos\bigl(2(\theta-\psi)\bigr)+
a\cos(\theta-\beta)\,.
\label{eqnum:e6}\end{equation}   

\vspace*{1pt}
Miller took this effect  into account  when analysing his data. The effect is apparent in Fig.~\ref{fig:Miller}, and even more so in the Michelson-Morley data in Fig.~\ref{fig:MM}.

\begin{figure}
\vspace*{.5mm}
\hspace{30mm}\includegraphics[scale=0.6]{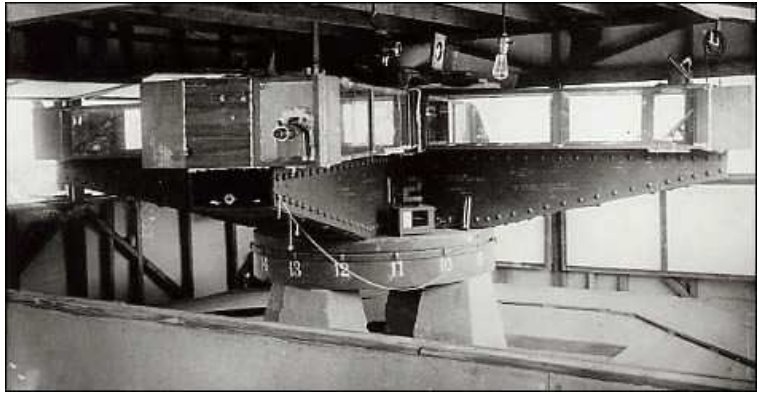}

\vspace*{-1.5mm}
\caption{\small{ Miller's interferometer with an effective arm length of $L\,{=}\,\mbox{32}$\,m achieved by multiple
reflections. Used by Miller on Mt.Wilson to perform the 1925-1926 observations of absolute motion. The steel arms weighed 1200 kilograms and floated in a tank of 275 kilograms of Mercury. From Case Western Reserve University Archives.}  
\label{fig:Millerinterf}}

\vspace*{-3mm}
\end{figure}
 
The interferometers are operated with the arms horizon\-tal, as shown by Miller's
interferometer in Fig.~\ref{fig:Millerinterf}. Then in (\ref{eqnum:e6}) $\theta$ is the azimuth of one arm
relative to the local meridian, while $\psi$ is the azimuth of the absolute motion
velocity projected onto the plane of the interferometer, with projected component $v_P$.
Here the Fitzgerald-Lorentz con\-traction is a real dynamical effect of absolute motion,
unlike the Einstein spacetime view that it is merely a spacetime perspective artifact, and
whose magnitude depends on the choice of observer. The instrument is operated by rotating
at a rate of one rotation over several minutes, and observing the shift in the fringe
pattern through a telescope during the rotation.  Then fringe shifts from six (Michelson
and Morley) or twenty (Miller) successive rotations are averaged to improve the signal to noise ratio, and the average sidereal time noted, giving  the Michelson-Morley data in Fig.~\ref{fig:MM}. or the Miller
data like that in Fig.~\ref{fig:Miller}. The form in (\ref{eqnum:e6}) is then fitted to such data by varying
the parameters $v_P$, $\psi$, $a$ and $\beta$,  The data from rotations is sufficiently clear, as in Fig.~\ref{fig:Miller}, that Miller could easily determine these parameters from a graphical plot. 

However Michelson and Morley implicitly assumed the
Newtonian value  $k{=}1$, while Miller used an indirect method to estimate the value of $k$,
as he understood that the New\-tonian theory was invalid, but had no other theory for the
interferometer.  Of course the Einstein postulates, as distinct from Special Relativity,  have that absolute motion has no meaning, and so effectively demands that $k{=}0$.  Using $k{=}1$  gives only a nominal value for $v_P$, being some 8--9\,km/s for the Michelson and Morley experiment, and some 10\,km/s from
Miller; \,the  difference  arising  from  the 
 different  latitu-

\noindent
des of Cleveland and Mt.\,Wilson, and from Michelson and Morley taking data at limited times.  So already Miller knew that his observations were consistent with those of Michel\-son and Morley, and so the  important need for reproduci\-bi\-lity was being confirmed.

\begin{figure}[t]
\hspace{-6mm}\includegraphics[scale=0.5]{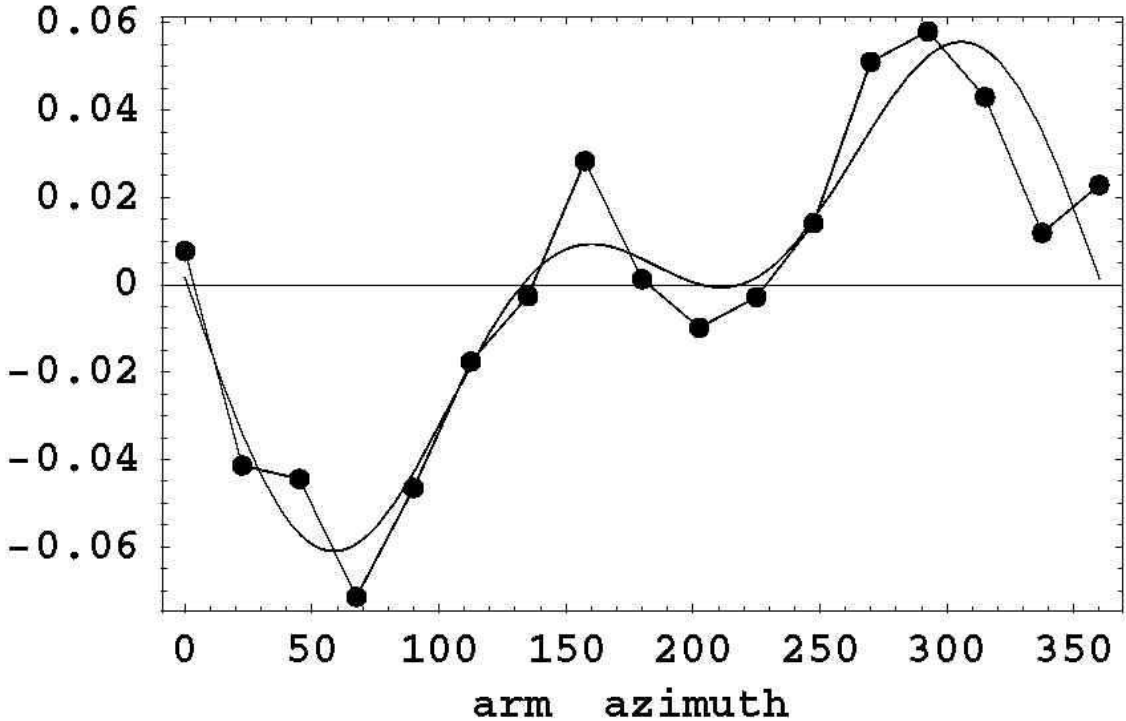}

\vspace*{-2.7mm}
\caption{\small Example of Michelson-Morley  fringe shifts from average of 6 rotations  measured every 22.5$^\circ$, in fractions of a wavelength $\Delta \lambda/\lambda$, vs  arm azimuth $\theta$(deg), from Cleveland, Ohio, July 11, 1887 12:00\,hrs local time or 7:00\,hrs  local sidereal time. This shows the quality of the fringe shift  data that Michelson and Morley obtained. The curve is the best fit using the form in (\ref{eqnum:e6}) which includes the  Hick's $\cos(\theta-\beta)$ component that is required when the mirrors are not orthognal, and gives
$\psi\,{=}\,\mbox{140}^\circ$, or 40$^\circ$  measured from South, compared to the Miller $\psi$ for August at 7:00\,hrs local sidereal time in Fig.~\ref{fig:MillerAz}, and a projected speed of $v_P=\mbox{400}$\,km/s.  The Hick's effect is much larger in this data than in the Miller data in  Fig.~\ref{fig:Miller}.}

\vspace*{-3.75mm}
\label{fig:MM}\end{figure}

\subsection{Michelson-Morley experiment}\label{subsect:32}

The Michelson and Morley air-mode interferometer fringe shift data was 
based upon a total of only 36 rotations in July 1887, revealing the nominal speed of some
8--9\,km/s when analysed using the prevailing but incorrect Newtonian theory which has  $k\,{=}\,1$ in 
(\ref{eqnum:e6}), and this value was known to Michelson and Morley. Including the Fitzgerald-Lorentz
dynamical contraction effect as well as the effect of the gas present as in (\ref{eqnum:e6})  we find that
 $n_{air}\,{=}\,\mbox{1.00029}$ gives $k^2\,{=}\,\mbox{0.00058}$  for air, which explains why the observed fringe shifts were so
small.  The example in Fig.~\ref{fig:MM} reveals a speed of 400\,km/s with an azimuth of 40$^\circ$ measured from south at 7:00 hrs local sidereal time.  The data is clearly very consistent with the expected form in (\ref{eqnum:e6}). They rejected their own data on the sole but spurious ground that the value of 8\,km/s was smaller than the speed of the Earth about the Sun of 30km/s.  What their result really showed was that (i) absolute motion had been detected because fringe shifts of
the correct form, as in  (\ref{eqnum:e6}), had been detected, and (ii) that the theory giving $k^2\,{=}\,1$  was
wrong, that Newtonian physics had failed. Michelson and Morley in 1887 should have announced that
the speed of light did depend of the direction of travel, that the speed  was relative to an
actual physical 3-space. However contrary to their own data they concluded that absolute
motion had not been detected.  This bungle has had enormous implications for fundamental
theo\-ries of space and time over the last 100 years, and the re\-sult\-ing confusion is only now
being finally corrected, albeit with fierce and spurious objections.

\begin{figure}[t]
\vspace*{-0mm}\includegraphics[scale=2.0]{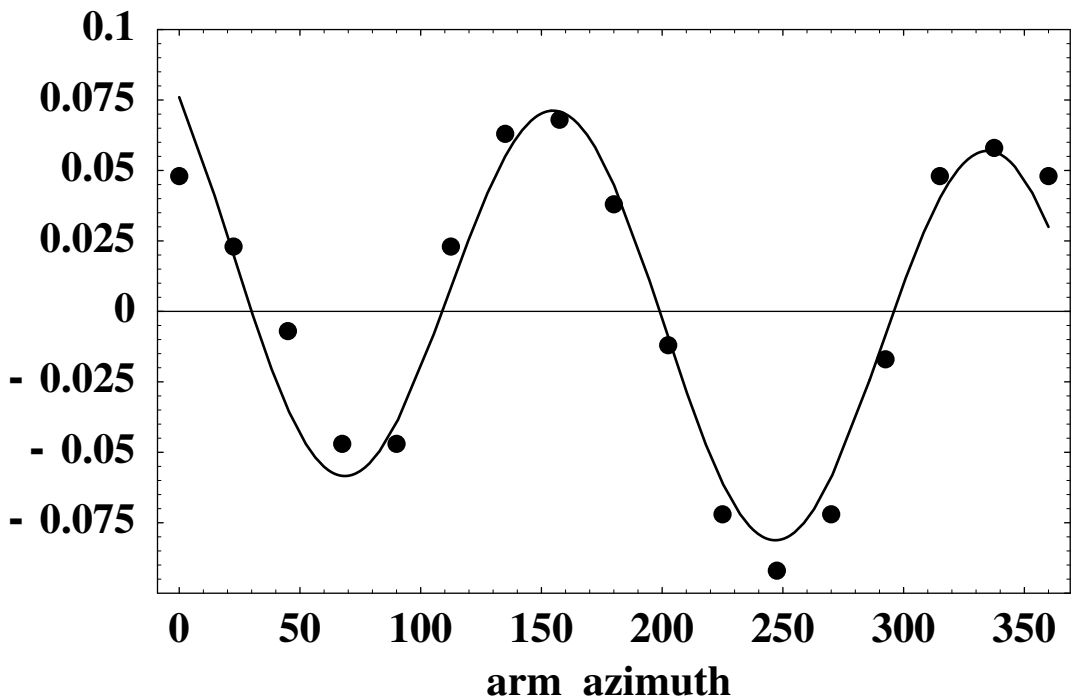} 

\vspace*{-2mm}
\caption{\small Typical Miller rotation-induced fringe shifts from average of 20 rotations,  measured every 22.5$^\circ$, in fractions of a wavelength $\Delta \lambda/\lambda$, vs  arm azimuth $\theta$(deg), measured
clockwise from North, from Cleveland Sept. 29, 1929 16:24 UT; 11:29\,hrs average local sidereal time. The curve is the best fit using the form in (\ref{eqnum:e6}) which includes the  Hick's $\cos(\theta-\beta)$ component that is required when the mirrors are not orthognal, and gives $\psi=\mbox{158}^\circ$, or $22^\circ$  measured from South, and a projected speed of $v_P =\mbox{351}$\,km/s.   This process was repeated some 8,000 times over days throughout 1925/1926 giving, in part, the data in Fig.~\ref{fig:MillerAz} and Fig.~\ref{fig:SeptPlot}.}
\label{fig:Miller}

\vspace*{-.5mm}
\end{figure}

\begin{figure}[t]
\vspace*{-13mm}
\vspace*{-0mm}\rule{-2mm}{0pt}\includegraphics[scale=2.0]{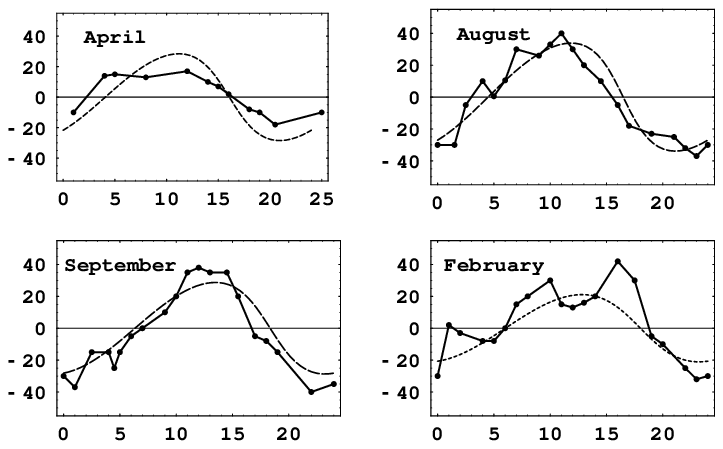}

\vspace*{-4mm}
\caption{\small  Miller azimuths $\psi$, measured from south and plotted against sidereal time in hours, showing both data and best fit of theory giving $v_{cosmic}\,{=}\,\mbox{433}$\,km/s in the direction 
($\mbox{RA}\,{=}\,\mbox{5.2}^{\mathrm{hr}}$, 
$\mbox{Dec}\,{=}{-}\mbox{67}^\circ$), and using $n\,{=}\,\mbox{1.000226}$ appropriate for the altitude of Mt.\,Wilson. The azimuth data gives a clearer signal than the speed data in Fig.~\ref{fig:SeptPlot}.  The data shows that the time when the azimuth $\psi$  is zero tracks sidereal time, with the zero times being approximately 5\,hrs and 17\,hrs.  However these times correspond to very different local times, for from April to  August, for example,  there is a shift of 8\,hrs in the local time for these crossings. This is an enormous effect. Again this is the acid test for light speed anisotropy experiments when allowing the rotation of the Earth to change the orientation of the apparatus.  The zero crossing times are when the velocity~vec\-tor for absolute motion when projected onto the plane of the~in\-ter\-fe\-ro\-meter lines up with the local meridian. As well we see~var\-i\-a\-tions throughout these composite days with the crossing times chang\-ing
by as much as  $\pm 3$\,hrs, The same effect, and perhaps even larger, is seen in the Flinders data in Fig.~\ref{fig:VaultData}. The above plots also show a distinctive signature, namely the change from month to month. This is caused by the vector addition of the Earth's orbital velocity of 30\,km/s, the Sun's spatial in-flow velocity of 42\,km/s at the Earth's dis\-tance  and the cosmic velocity changing over a year.  This is the~ef\-fect that Miller used to calibrate his interferometer. However he did not know of the Sun in-flow component. Only after taking ac\-count of that effect does this calibration method agree with the~results from the calibration method using Special Relativity, as~in~(\ref{eqnum:e6}).}  
\label{fig:MillerAz}

\vspace*{-1.5mm}
\end{figure}

\vspace*{2pt}
\subsection{Miller interferometer}\label{subsect:33}

\vspace*{-1pt}
It was Miller \cite{C4} who saw the flaw in the 1887 paper and realised that the theory for
the Michelson interferometer must be wrong.  To avoid using that theory Miller introduced
the scaling factor $k$, even though he had no theory for its value. He then used the effect of
the changing vector addition of the Earth's orbital velocity and the absolute galactic
vel\-oc\-ity of the solar system to determine the numerical value of  $k$, because the orbital
motion modulated the data, as shown in Fig.~\ref{fig:MillerAz}. By making some 8,000 rotations of the
interferometer at Mt.\,Wilson in 1925/26 Miller determined the first estimate for $k$  and for
the absolute linear velocity of the solar system. Fig.~\ref{fig:Miller} shows typical data from averaging
the fringe shifts from 20 rotations of the Miller interferometer, performed over a short
period of time, and clearly shows the expected form in (\ref{eqnum:e6}) (only a linear drift caused by
temperature effects on the arm lengths has been removed --- an effect also re\-mov\-ed by
Michelson and Morley and also by Miller). In Fig.~\ref{fig:Miller} the fringe shifts during rotation are
given as fractions of a wavelength, $\Delta \lambda/\lambda\,{=}\,\Delta t/T$, where $\Delta t$ is given by 
(\ref{eqnum:e6}) and $T$  is the period of the light. Such rotation-induced fringe shifts clearly show that
the speed of light is different in dif\-ferent directions. The claim that Michelson interferometers,
operating in gas-mode, do not produce fringe shifts under rotation is clearly incorrect. But it is that
claim that lead to the continuing belief, within physics, that absolute motion had never been detected,
and that the speed of light is invar\-iant. The value of $\psi$  from such rotations together lead
to plots like those in Fig.~\ref{fig:MillerAz}, which show $\psi$ from the 1925/1926 Miller \cite{C4}
interferometer data for four different months of the year, from which the RA\,$=$\,5.2\,hr is
readily apparent. While the orbital motion of the Earth about the Sun slightly affects the
RA in each month, and Miller used this effect do determine the value of  $k$, the new theory of
gravity required a reanalysis of the data \cite{C11,C9}, revealing that the solar system has
a large observed galactic velocity of some 420$\pm$30\,km/s in the direction (RA\,$=$\,5.2\,hr, Dec\,${=}{-}$67$^\circ$). This is different from the speed of 369\,km/s in the direction (RA\,${=}$\,11.20\,hr, Dec\,${=}{-}$7.22$^\circ$) extracted from the Cosmic Microwave Background (CMB) anisotropy, and which
de\-scribes a motion relative to the dis\-tant universe, but not relative to the local 3-space.
The Miller velocity is explained by galactic gravitational in-flows~\cite{C11}.

\vspace*{-3pt}
\subsection{Other gas-mode Michelson interferometer 
experiments}\label{subsect:34}

\vspace*{-3pt}
Two old interferometer experiments, by Illingworth \cite{C5} and Joos \cite{C6}, used
helium, enabling the refractive index effect to be recently confirmed, because for helium,
with $n\,{=}$ ${=}\,\mbox{1.000036}$, we find that 
$k^2\,{=}\,\mbox{0.00007}$.  Until the refractive index effect was taken into account
the data from the helium-mode experiments appeared to be inconsistent with the data from the air-mode
experiments; now they are seen to be consistent \cite{C11}. Ironically helium was introduced in place of air to
reduce any possible unwanted effects of a gas, but we now understand the essential role of
the gas. The data from an interferometer experiment by Jaseja {\it  et al.}
\cite{C7}, using two
orthogonal masers with a He-Ne gas mixture, also indicates that they detected absolute
motion, but were not aware of that as they used the incorrect Newtonian theory and so
considered the fringe shifts to be too small to be real, reminiscent of the same mistake by
Michelson and Morley. The Michelson interferometer is a 2nd order device, as the effect of
absolute motion is proportional to  $(v/c)^2$, as in  (\ref{eqnum:e6}), but 1st order devices are also possible
and the coaxial cable experiments described next are in this class. The experimental results and the implications for physics have been extensively reported in \cite{C11,MMC,AIP,NPA,IE,EP}.

\vspace*{-3pt}
\subsection{Coaxial cable speed of EM waves  anisotropy 
experiments}\label{subsect:35}

\vspace*{-3pt}
Rather than use light travel time experiments to demonstrate the anisotropy of the speed of light another technique is to measure the one-way speed of radio waves through a coaxial electrical cable.  While this not a direct ``ideal'' technique, as then the complexity of the propagation physics comes into play, it provides not only an independent confirmation of the light anisotropy effect, but also one which takes advantage of modern electronic timing technology.

\vspace*{-3pt}
\subsection{Torr-Kolen coaxial cable anisotropy  experiment}\label{subsect:36}

\begin{figure}[h]
\hspace{30mm}\includegraphics[scale=1.15]{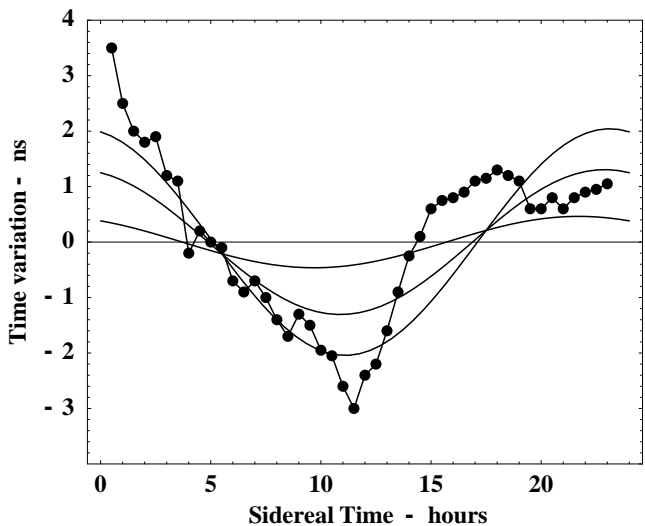}

\caption{\small{ 
Data from one day of the Torr-Kolen EW coaxial cable an\-iso\-tropy experiment.  Smooth curves show variations in travel times when the declination is varied by ${\pm}\,\mbox{10}^\circ$ about the direction ($\mbox{RA}\,{=}\,\mbox{5.2}^{\mathrm{hr}}, \mbox{Dec}\,{=}{-}\mbox{67}^\circ$), for a cosmic speed of 433\,km/s.  Most impor\-tant\-ly the dominant feature is consistent with the predicted local sidereal time.}
\label{fig:TK}}

\vspace*{1.9mm}
\end{figure}

\vspace*{-2pt}
The first one-way coaxial cable speed-of-propagation exper\-i\-ment was performed at the Utah University in 1981 by Torr and Kolen. This involved two rubidium  clocks placed approximately 500\,m apart with a 5\,MHz radio frequency (RF) signal propagating between the clocks via a buried EW nitrogen-filled coaxial cable maintained at a constant pres\-sure of 2\,psi. Torr and Kolen  found that, while the round speed time remained constant within 0.0001\%\,$c$, as expected from Sect.~\ref{sect:SR}, variations in the one-way travel time  were observed. The maximum effect occurred, typically, at the times predicted using the Miller galactic velocity, although Torr and Kolen appear to have been unaware of the Miller experiment. As well Torr and Kolen reported fluctuations in both the magnitude, from 1--3\,ns, and the time of maximum variations in travel time.  These effects are interpreted as arising from the turbulence in the flow of space past the Earth.  One day of their data is shown in Fig.~\ref{fig:TK}.

\begin{figure}[t]
\hspace{25mm}\includegraphics[scale=1.4]{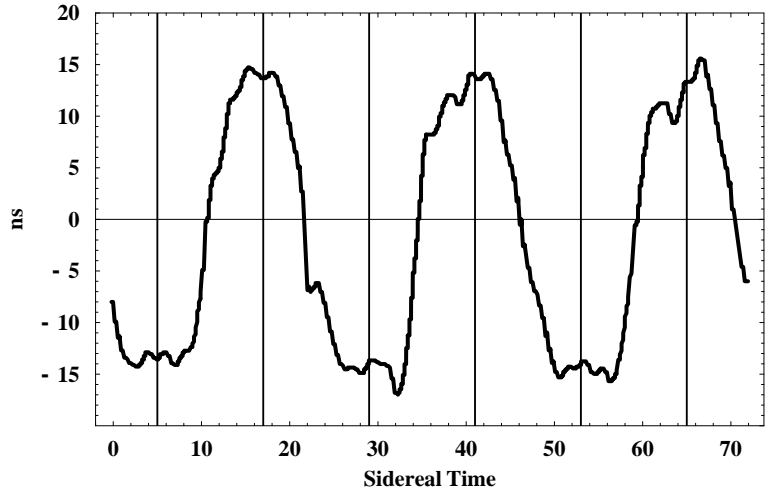}

\vspace*{-3.0mm}
\caption{\small{ Variations in twice the one-way travel time, in ns, for an RF signal to travel 1.5\,km through a coaxial cable between  Rue du Marais and Rue de la Paille, Brussels. 
An offset  has been used  such that the average is zero.   The cable has a
North-South  orientation, and the data is $\pm$ difference of the travel times  for NS and SN
propagation.  The sidereal time for maximum  effect of ${\sim}\,$5\,hr and   ${\sim}\,$17\,hr (indicated
by vertical lines) agrees with the direction found by Miller. Plot shows
data over 3 sidereal days  and is plotted against sidereal time. 
 The fluctuations are evidence of turbulence  of gravitational waves.}  
\label{fig:DeWittetimes}}

\vspace*{-2.0mm}
\end{figure}

\vspace*{2pt}
\subsection{De Witte coaxial cable anisotropy experiment}\label{subsect:37}

\vspace*{1pt}
During 1991 Roland De Witte performed a most extensive RF coaxial cable travel-time anisotropy experiment, accu\-mu\-lating data over 178 days. His data is in complete agreement with the Michelson-Morley 1887 and  Miller 1925/26 inter\-ferometer experiments.   The Miller and De Witte experiments will eventually be recognised as two of the most significant experiments in physics, for independently and using different experimental techniques they detected essentially the same velocity of absolute motion.  But also they detected turbu\-lence in the flow of space past the Earth --- none other than gravitational waves.
The De Witte experiment was within Belgacom, the Belgium telecommunications company. This organisation had two sets of atomic clocks in two buildings in Brussels separated by 1.5\,km and the research project was an investigation of  the task of synchronising these two clusters of atomic clocks. To that end 5MHz RF signals were sent in both directions   through two buried coaxial cables linking the two clusters.   The atomic clocks were caesium beam atomic clocks, and there were three in each cluster: A1, A2 and A3 in one cluster, and B1, B2, and B3 at the other cluster. In that way the stability of the clocks could be established and monitored. One cluster was in a building on Rue du Marais and the second cluster was due south in a building on Rue de la Paille.  Digital phase comparators were used to measure changes in times between clocks within the same cluster and also in the one-way propagation times of the RF signals.  At both locations the comparison between local clocks, A1-A2 and A1-A3, and between B1-B2, B1-B3, yielded linear phase variations in agreement with the fact that the clocks have not exactly the same frequencies together with a short term and long term phase noise. But between distant clocks A1 toward B1 and B1 toward A1, in addition to the same linear phase variations, there is also an additional clear sinusoidal-like phase undulation with an ap\-proximate 24\,hr period of the order of 28\,ns peak to peak, as shown in Fig.~\ref{fig:DeWittetimes}. The possible instability of the coaxial lines cannot be responsible for the observed phase effects because these signals are in phase opposition and also because the lines are identical (same place, length, temperature, etc\dots) causing the cancellation of any such instabilities. As well the experiment was performed over 178 days, making it possible to measure with an accuracy of  25\,s the period of the phase signal to be the sidereal day (23\,hr 56\,min).

\begin{figure}[t]
\hspace{40mm}\includegraphics[scale=1.0]{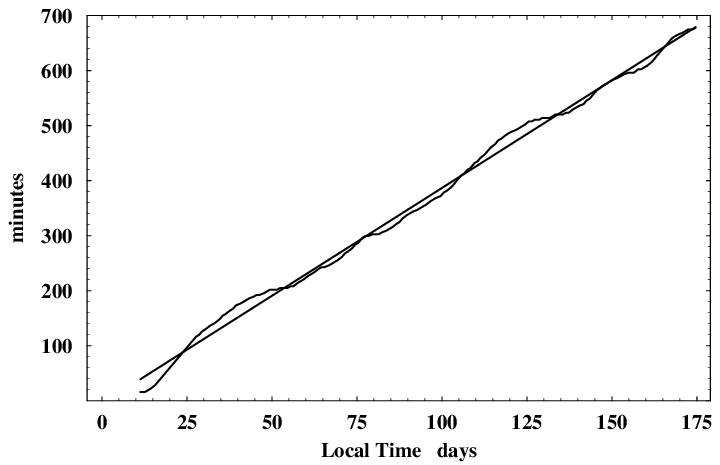}

\vspace*{2mm}
\hspace{40mm}\includegraphics[scale=0.23]{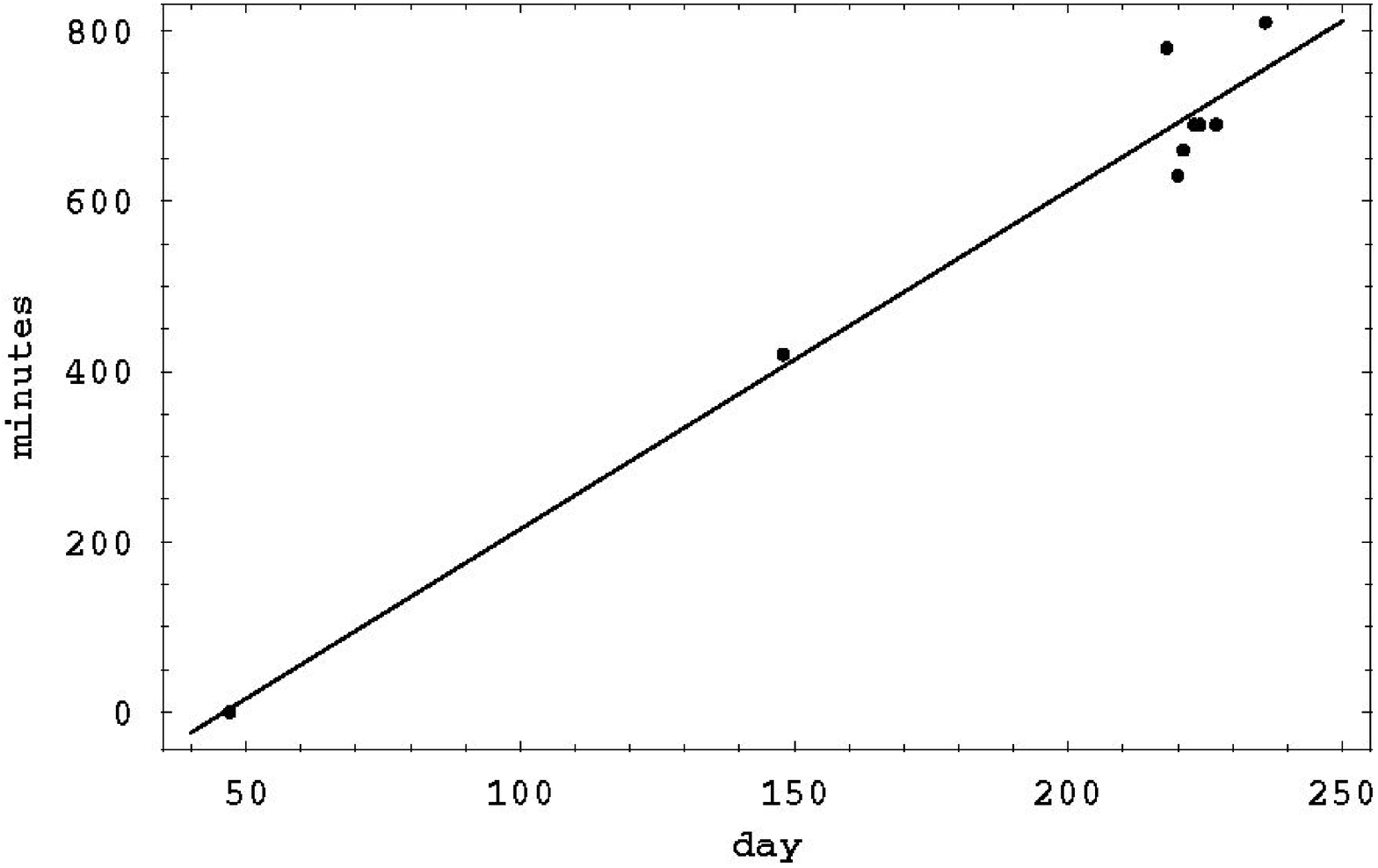}

\vspace*{-2.5mm}
\caption{\small{ {\bf Upper:} Plot from the De Witte data of the negative of the drift of the cross-over time  between minimum and
maximum travel-time variation each day (at ${\sim}\,\mbox{10hr}
{\pm}\mbox{1hr}$\,ST) versus local solar time for some
180 days. The straight line plot is the least-squares fit to the experimental data, 
 giving an average slope of 3.92 minutes/day. The time difference between a sidereal day and a solar
day is 3.93 minutes/day.    This demonstrates that the effect is related to sidereal time and not local
solar time.   {\bf Lower:} Analogous sidereal effect seen in the Flinders experiment. Due to on-going developments the data is not available for all days, but sufficient data is present to indicate a  time shift of 3.97 minutes/day. This data also shows greater fluctuations than indicated by the De Witte data, presumably because De Witte used more extensive data averaging.}  
\label{fig:DeWitteST}}

\vspace*{-1.9mm}
\end{figure}

\begin{figure}
\hspace{15mm}\includegraphics[scale=0.4]{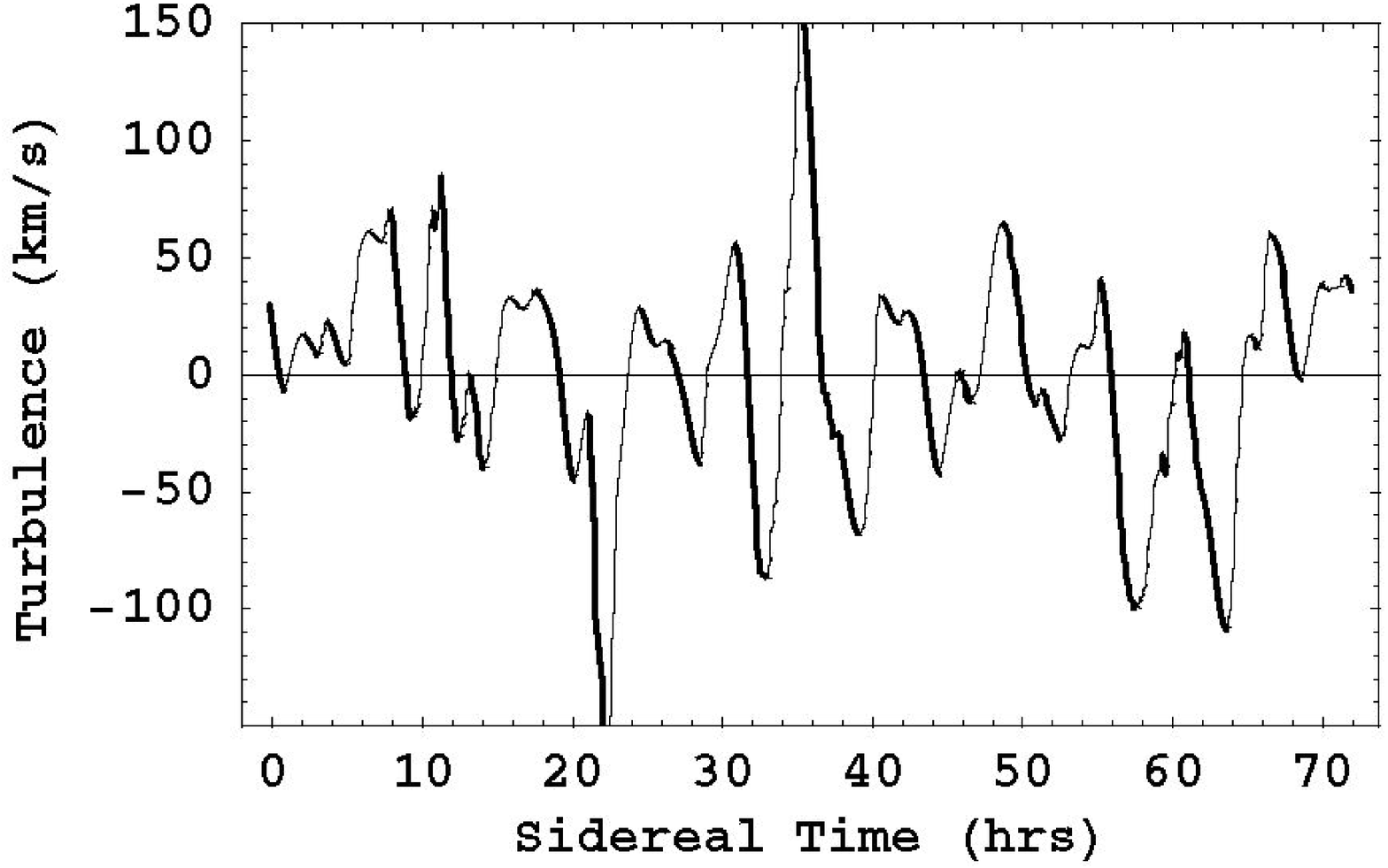}

\vspace*{-3.0mm}
\caption{\small{Shows the speed fluctuations, essentially ``gravitational waves'' observed by      De Witte in 1991 from the measurement of variations in the RF coaxial-cable travel times.  This data is obtained from that in Fig.~\ref{fig:DeWittetimes} after removal of the dominant effect caused by the rotation of the Earth.  Ideally the velocity fluctuations are three-dimensional, but the De Witte experiment had only one  arm. This plot is suggestive of a fractal structure to the velocity field. This is confirmed by the power law analysis   in  \cite{Schrod, DeWitte}.}
\label{fig:fractal}}

\vspace*{-3.5mm}
\end{figure}

Changes in propagation times were observed over 178 days from June 3 to November 27, 1991. A sample of the data, plotted against sidereal time for just three days, is shown in Fig.~\ref{fig:DeWittetimes}.  De Witte recognised that the data was evi\-dence of absolute motion but he was unaware of the Miller experiment and did not realise that the Right Ascensions for minimum/maximum propagation time agreed almost exactly with that predicted using the Miller's direction (RA$\,{=}\,$5.2\,hr, Dec$\,{=}{-}$67$^\circ$). In fact De Witte expected that the direction of 
absolute motion should have been in the CMB direction, but that would have given the data a totally different sidereal time signature, namely the times for maximum/minimum would have been shifted by 6\,hrs.  The declination of the velocity observed in this De Witte experiment cannot be de\-t\-er\-mined from the data as only three days of data are avai\-lable.   The De Witte data is analysed in Sect.~\ref{sect:gravity} and as\-sum\-ing a declination of 60$^\circ$\,S a speed of 430\,km/s is obtained, in good agreement with the Miller speed and Michelson-Morley speed. So a different and non-relativistic technique is con\-firm\-ing the results of these older experiments. This is dramatic.

De Witte did however report the sidereal time of the cross-over time, that is in Fig.~\ref{fig:DeWittetimes} for all 178 days of data.  That showed, as in Fig.~\ref{fig:DeWitteST}, that the time variations are correlated with sidereal time and not local solar time. A least-squares best fit of a linear relation to that data gives that the cross-over time is retarded, on average, by 3.92 minutes per solar day. This is to be compared with the fact that a sidereal day is 3.93 minutes shorter than a solar day. So the effect is certainly galactic and not associated with any daily thermal effects, which in any case would be very small as the cable is buried.  Miller had also compared his data against sidereal time and established the same property, namely that the diurnal effects  actually tracked sidereal time and not solar time,  and that orbital effects were also apparent, with both effects apparent   in Fig.~\ref{fig:MillerAz}.

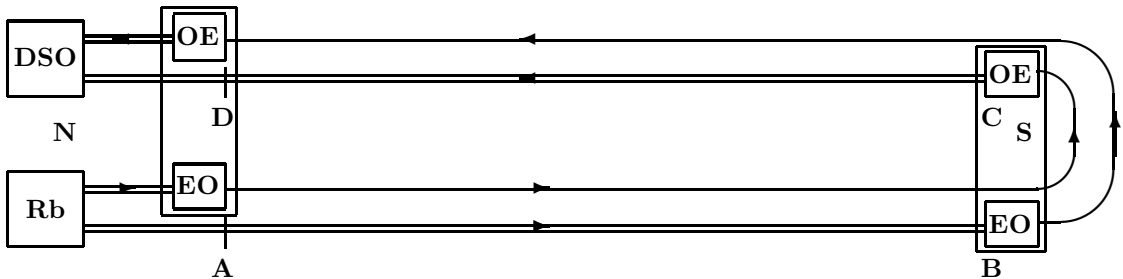
\begin{figure*}[ht]
\setlength{\unitlength}{2.0mm}
\vspace{30mm}
\hspace{12mm}\begin{picture}(0,0)
\thicklines
\put(0,0){\line(0,1){5}}\put(0,0){\line(1,0){5}}
\put(0,5){\line(1,0){5}}\put(5,0){\line(0,1){5}}
\put(1.2,1.9){\bf Rb}

\put(0,10){\line(0,1){5}}\put(0,10){\line(1,0){5}}
\put(0,15){\line(1,0){5}}\put(5,10){\line(0,1){5}}
\put(0.4,12){\bf DSO}

\put(5,1){\line(1,0){60}}
\put(5,1.4){\line(1,0){60}}
\put(35,1.35){\vector(1,0){1}}

\put(5,4){\line(1,0){6}}
\put(5,3.6){\line(1,0){6}}

\put(5,11){\line(1,0){60}}
\put(5,11.4){\line(1,0){60}}

\put(5,14){\line(1,0){6}}
\put(5,13.6){\line(1,0){6}}

\put(35,1.35){\vector(1,0){1}}
\put(35,13.8){\vector(-1,0){1}}
\put(8,13.8){\vector(-1,0){1}}
\put(35,11.2){\vector(-1,0){1}}
\put(35,3.9){\vector(1,0){1}}

\put(7.5,3.9){\vector(1,0){1}}


\put(11,2.5){\line(0,1){3}}
\put(11,2.5){\line(1,0){3.5}}
\put(11,5.5){\line(1,0){3.5}}
\put(14.5,2.5){\line(0,1){3}}
\put(11.2,3.5){\bf EO}

\put(13.5,-2){\bf A}
\put(14.5,-0.1){\line(0,1)2}
\put(64.7,-2){\bf B}

\put(13.5,8){\bf D}
\put(14.5,9.9){\line(0,1)2}
\put(64.7,8){\bf C}

\put(65,0.0){\line(0,1){3}}
\put(65,0.0){\line(1,0){3.5}}
\put(65,3.0){\line(1,0){3.5}}
\put(68.5,0){\line(0,1){3}}
\put(65.2,0.9){\bf EO}

\put(65,10.0){\line(0,1){3}}
\put(65,10.0){\line(1,0){3.5}}
\put(65,13.0){\line(1,0){3.5}}
\put(68.5,10){\line(0,1){3}}
\put(65.2,10.9){\bf OE}

\put(11,12.4){\line(0,1){3}}
\put(11,12.4){\line(1,0){3.5}}
\put(11,15.5){\line(1,0){3.5}}
\put(14.5,12.4){\line(0,1){3}}
\put(11.2,13.4){\bf OE}

\put(14.5,13.7){\line(1,0){54}}
\put(14.5,3.8){\line(1,0){54}}
\put(68.4,8.92){\oval(5,10.1)[rb]}
\put(68.4,6.6){\oval(5,10.1)[rt]}

\put(68.5,6.67){\oval(10,10.1)[rb]}
\put(68.5,8.68){\oval(10,10)[rt]}
\put(73.5,6){\line(0,1){4}}
\put(73.6,6){\vector(0,1){3}}
\put(70.9,6.0){\vector(0,1){2}}

\put(3,7){\bf N}\put(67,7){\bf S}

\put(10.2,2.1){\line(0,1){13.8}}
\put(10.2,2.1){\line(1,0){5.0}}
\put(10.2,15.9){\line(1,0){5.0}}
\put(15.2,2.1){\line(0,1){13.8}}

\put(64.4,-0.33){\line(0,1){13.6}}
\put(64.4,-0.33){\line(1,0){4.6}}
\put(64.4,13.32){\line(1,0){4.6}}
\put(69.0,-0.33){\line(0,1){13.6}}

\end{picture}
\vspace*{1.5mm}\caption{Schematic layout of the Flinders University Gravitational Wave Detector. Double lines denote coaxial cables, and single lines denote optical fibres.  The detector is shown in Fig.~\ref{fig:Expta} and is orientated NS along the local meridian, as indicated by direction D in Fig.~16. Two 10\,MHz RF signals come from the Rubidium atomic clock (Rb).  The Electrical to Optical converters (EO) use the RF signals to modulate 1.3\,$\mu$m infrared signals that propagate through the single-mode optical fibres.  The Optical to Electrical converters (OE) demodulate that signal and give the two RF signals that finally reach the Digital Storage Oscilloscope (DSO), which measures their phase difference.  Pairs of E/O and O/E are grouped into one box. Overall this apparatus measures the {\it difference} in EM travel time from A to B compared to C to D. All other travel times cancel in principle, though in practice small differences in cable or fibre lengths need to be electronically detected by the looping procedure. The key effects are that the propagation speeds through the coaxial cables and optical fibres respond differently to their absolute motion through space.  The special optical fibre propagation effect is discussed in the text.  Sections AB and CD each have length 5.0\,m.  The fibres and coaxial cable are specially manufactured to have negligible variation in travel speed with variation in temperature.   The zero-speed calibration point can be measured by looping the arm back onto itself, as shown in Fig.~\ref{fig:Exptb}, because then the 1st order in $v/c$ effect cancels, and only 2nd order effects remain, and these are much smaller than the noise levels in the system.  This detector is equivalent to a one-way speed measurement through a single coaxial cable of length 10\,m, with an atomic clock at each end to measure changes in travel times.  However for 10\,m coaxial cable that would be impractical because of clock drifts. With this set-up the travel times vary by some 25\,ps over one day, as shown in Figs.\ref{fig:Loop} and \ref{fig:STEffect}.  The detector was originally located in the author's office, as shown in Fig.~\ref{fig:Expta}, but was later located in an underground laboratory where temperature variations were very slow.   The travel time variations over 7 days are shown in Fig.~\ref{fig:VaultData}.
\label{fig:FUGWD}}

\vspace*{-3.0mm}
\end{figure*}

The dominant effect in Fig.~\ref{fig:DeWittetimes}  is caused by the rotation of the Earth, namely that the orientation of the coaxial cable with respect to the average direction of the flow past the Earth changes as the Earth rotates. This effect may be ap\-proximately unfolded from the data leaving the gravitational waves shown in Fig.~\ref{fig:fractal}.  This is the first evidence that the velocity field describing the flow of space has a complex structure, and is indeed fractal. The fractal structure, i.\,e. that there is an intrinsic lack of scale to these speed fluctuations, is demonstrated by binning the absolute speeds   and counting the number of speeds  within each bin,  as discussed in \cite{Schrod, DeWitte}. The Miller data also shows evidence of turbulence of the same magnitude.  So far the data from three experiments, namely Miller, Torr and Kolen, and De Witte, show turbu\-len\-ce in the flow of space past the Earth.  This is what can be called gravitational waves.  This can be understood by noting that fluctuations in the velocity field induce ripples in the mathematical construct known as spacetime, as in (\ref{eqnum:E14}).  Such ripples in spacetime are known as gravitational waves.


\markright{R.\,T.\,Cahill. A New  Light-Speed Anisotropy Experiment: Absolute
Motion and Gravitational Waves Detected}

\section{Flinders University gravitational wave detector}\label{sect:4}

\markright{R.\,T.\,Cahill. A New  Light-Speed Anisotropy Experiment: Absolute
Motion and Gravitational Waves Detected}

\vspace*{-2pt}
In February 2006 first measurements from a gravitational wave  detector  at  Flinders  University,  Adelaide,  were  taken.

\noindent
This detector uses a novel timing scheme that overcomes the limitations associated with the two previous coaxial cable experiments.  The intention in such experiments is simply to measure the one-way travel time of RF waves propagating through the coaxial cable.  To that end one would apparently require two very accurate clocks at each end, and associated RF generation and detection electronics.

However the major limitation is that even the best atomic clocks are not sufficiently accurate over even a day to make such \rule{-.5pt}{0pt}measurements \rule{-.5pt}{0pt}to \rule{-.5pt}{0pt}the \rule{-.5pt}{0pt}required \rule{-.5pt}{0pt}accuracy, \rule{-.5pt}{0pt}unless \rule{-.5pt}{0pt}the \rule{-.5pt}{0pt}cables are of order of a kilometre or so in length, and then tem\-pe\-ra\-ture control becomes a major problem. The issue is that the time variations are of the order of 25\,ps per 10 meters of cable. To measure that requires time measurements accurate to, say,  1\,ps.  But atomic clocks have accuracies over one day of around 100\,ps, implying that lengths of around 1 kilometre would be required, in order for the effect to well exceed timing errors. Even then the atomic clocks must be brought together every day to resynchronise them, or use De Witte's method of multiple atomic clocks.  However at Flinders University a major breakthrough for this problem was made when it was discovered that unlike coaxial cables, the movement of optical fibres through space does not affect the propagation speed of light through them.  This is a very strange effect and at present there is no explanation for it.

\begin{figure*}[t] 
\vspace{0mm}
\parbox{69mm}{\vspace*{-18.5mm}\caption{The Flinders University Grav\-i\-ta\-tional Wave Detector located in the author's office, showing the Rb atomic clock and Digital Storage Oscilloscope (DSO) at the Northern end of the NS 5\,m cable run. In the foreground is  one Fibre Optic Trans\-ceiver. The coax\-ial cables are black, while the optical fibres are tied together in a white plastic sleeve, except just prior to con\-nect\-ing with the transceiver.  The second pho\-to\-graph shows the other transceiver at the Southern  end.\protect  Most of the data reported herein was taken when the de\-tect\-or was reloc\-at\-ed to an isolated under\-ground lab\-o\-ra\-tory with the transceivers resting on a con\-crete floor for temperature stabilisation.\label{fig:Expta}}}
\rule{1mm}{0pt}\parbox{104mm}{\hspace{4.5mm}\includegraphics[scale=0.115]{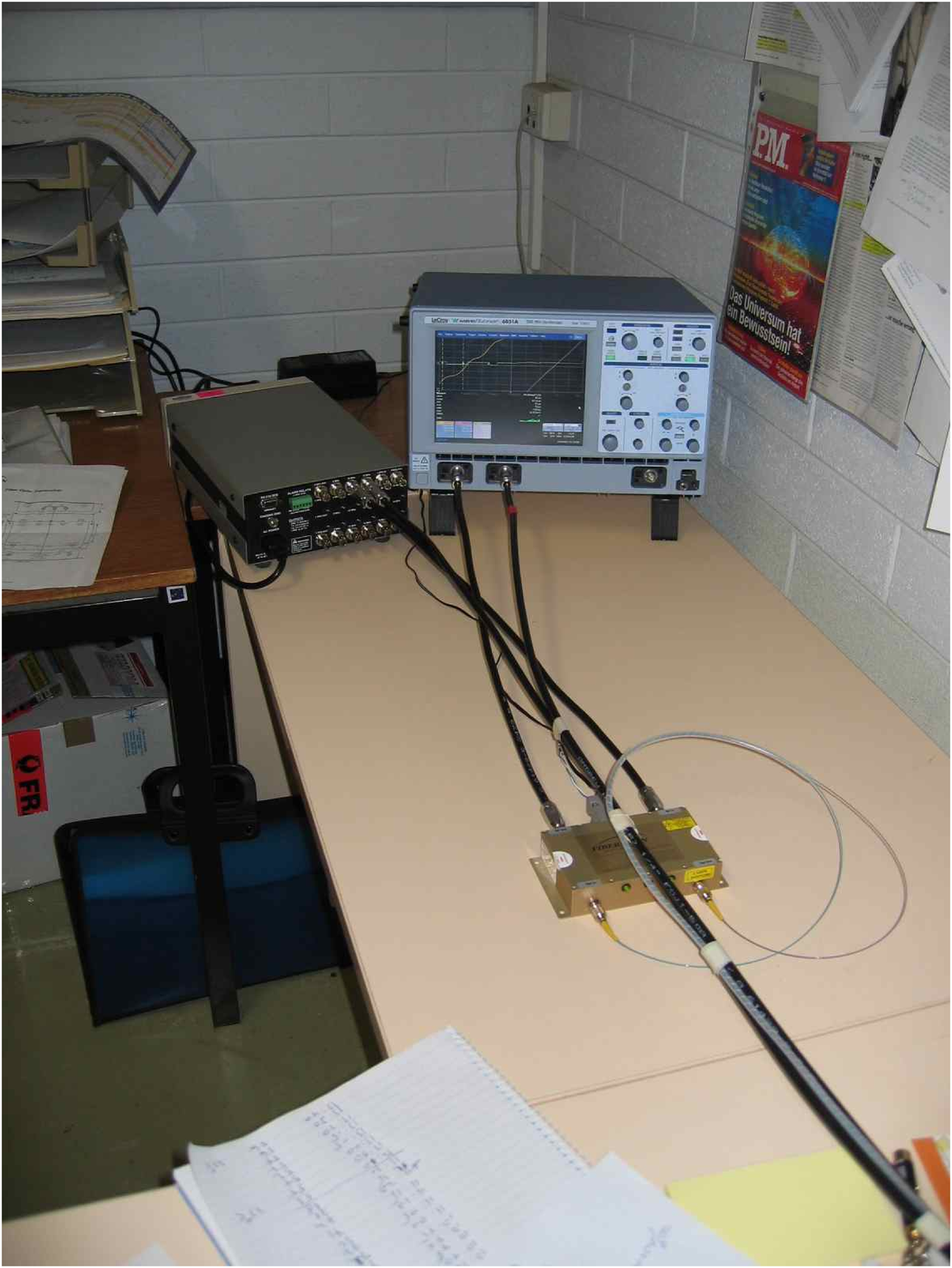}
\rule{1.5mm}{0pt}\includegraphics[angle=0,scale=.115]{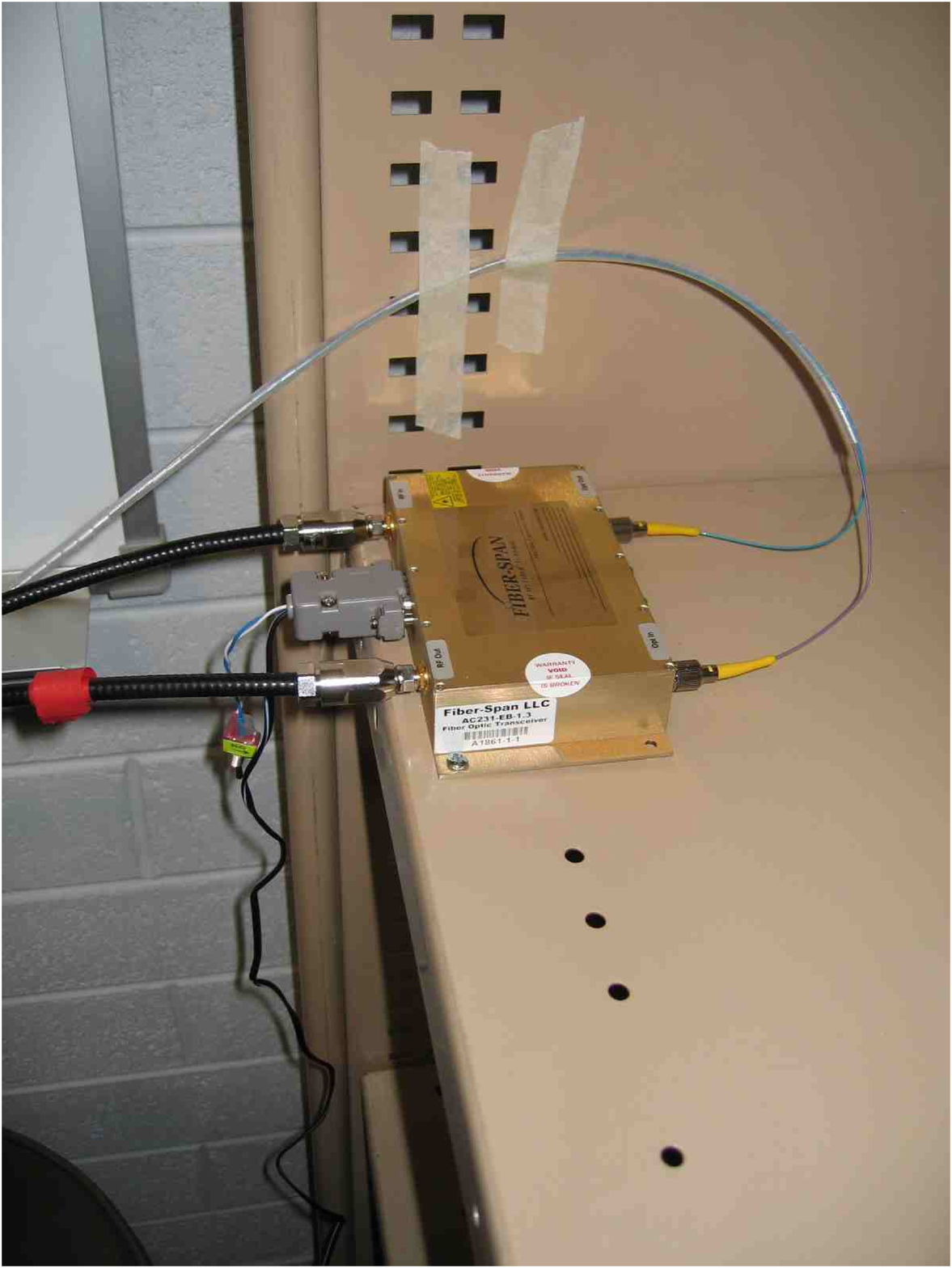}}

\vspace*{-4.8mm}
\end{figure*}

\vspace*{2pt}
\subsection{Optical fibre effect}\label{subsect:41}

\vspace*{1pt}
This effect was discovered by Lawrance, Drury and the author, using optical fibres in a Michelson interferometer 
arrangement, where the effective path length in each arm was 4 metres of fibre. So rather than having light pass through a gas, and being reflected by mirrors, here the light propagates through fibres and, where the mirrors would normally be located, a 180 degree bend in the fibres is formed. The light emerging from the two fibres is directed to a common region on a screen, and the expected fringe shifts were seen.  However, and most dramatically, when the whole apparatus was rotated no shift in the fringe shifts was seen, unlike the situation with light passing through a gas as above.   This result implied that the travel time in each arm of the fibre was unaffected by the orientation of that arm to the direction of the spatial flow. While no explanation has been developed for this effect, other than the general observation that the propagation speed in optical fibres depends on refractive index profiles and transverse and longitudinal Lorentz con\-traction effects, as in solids these are coupled by the elastic properties of the solid.  Nevertheless this property offered a technological leap forward in the construction of a compact coaxial cable gravitational wave detector.  This is because timing information can be sent though the fibres in a way that is not affected by the orientation of the fibres, while the coaxial cables do respond to the anisotropy of the speed of EM radiation in vacuum. Again why they respond in this way is not understood.  All we have is that fibres and coaxial cables respond differently.   So this offers the opportunity to have a coaxial cable one-way speed measurement set up, but using only one clock, as shown in Fig.~\ref{fig:FUGWD}.   Here we have one clock at one end of the coaxial cable, and the arrival time of the RF signal at the other end is used to modulate a light signal that returns to the starting end via an optical fibre. The return travel time is constant, being independent of the orientation of the detector arm, because of this peculiar property of the fibres.  In practice one uses two such arrangements, with the RF directions opposing one another.  This has two significant advantages, (i) that the effective coaxial cable length of 10 meters is achieved over a distance of just 5 meters, so the device is more easily ac\-com\-modated in a temperature controlled room,  and (ii) tem\-pe\-ra\-ture variations in that room have a smaller effect than expected because it is only temperature differences between the cables that have any net effect.  Indeed with specially con\-structed phase compensated fibre and coaxial cable, having very \rule{-.3pt}{0pt}low \rule{-.3pt}{0pt}speed-sensitivity \rule{-.3pt}{0pt}to \rule{-.3pt}{0pt}temperature \rule{-.3pt}{0pt}variations, the \rule{-.3pt}{0pt}most temperature sensitive components are the optical fibre trans\-ceivers (E/O and O/E in Fig.~\ref{fig:FUGWD}).

\subsection{Experimental components}\label{subsect:42}

\noindent {\bf Rubidium Atomic Clock:}
Stanford Research System  FS725 Rubidium Frequency Standard. Multiple 10MHz RF outputs. Different outputs were used for the two circuits of the detector.

\smallskip\noindent {\bf Digital Storage Oscilloscope:}
LeCroy WaveRunner WR6051A 500\,MHz 2-channel Digital Storage Oscilloscope (DSO).
Jitter Noise Floor 2\,ps rms. Clock Accuracy <5\,pm. DSO averaging set at 5000, and  generating time readings at 440/minute. Further averaged in DSO  over 60 seconds, giv\-ing stored data stream at one data point/minute. The data was further running-averaged over a 60 minute interval.  Con\-necting the Rb clock directly to the DSO via its two channels showed a long-term accuracy of $\pm 1$\,ps  rms with this setup.

\begin{figure}[t]
\begin{center}
\vspace*{1mm}
\includegraphics[scale=0.13]{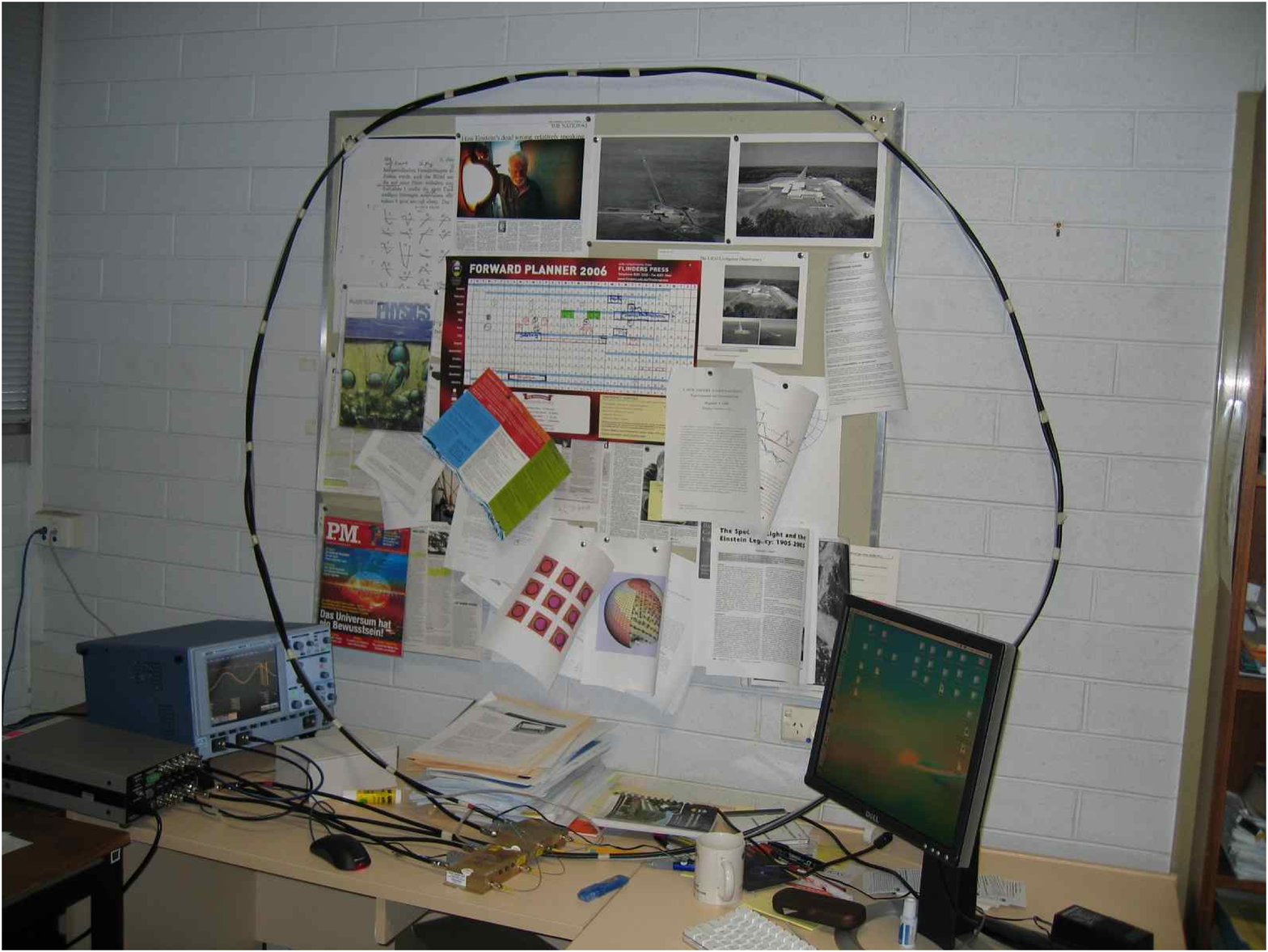}
\end{center}

\vspace*{-5mm}
\caption{The Flinders University Gravitational Wave Detector showing the cables formed into a loop. This configuration enables the calibration of the detector.  The data from such a looping is shown in Fig.~\ref{fig:Loop}, but  when the detector was relocated to an isolated underground laboratory.}
\label{fig:Exptb}

\vspace*{-2.5mm}
\end{figure}

\begin{figure}[t]
\hspace*{30mm}\hspace{5mm}\includegraphics[scale=0.25]{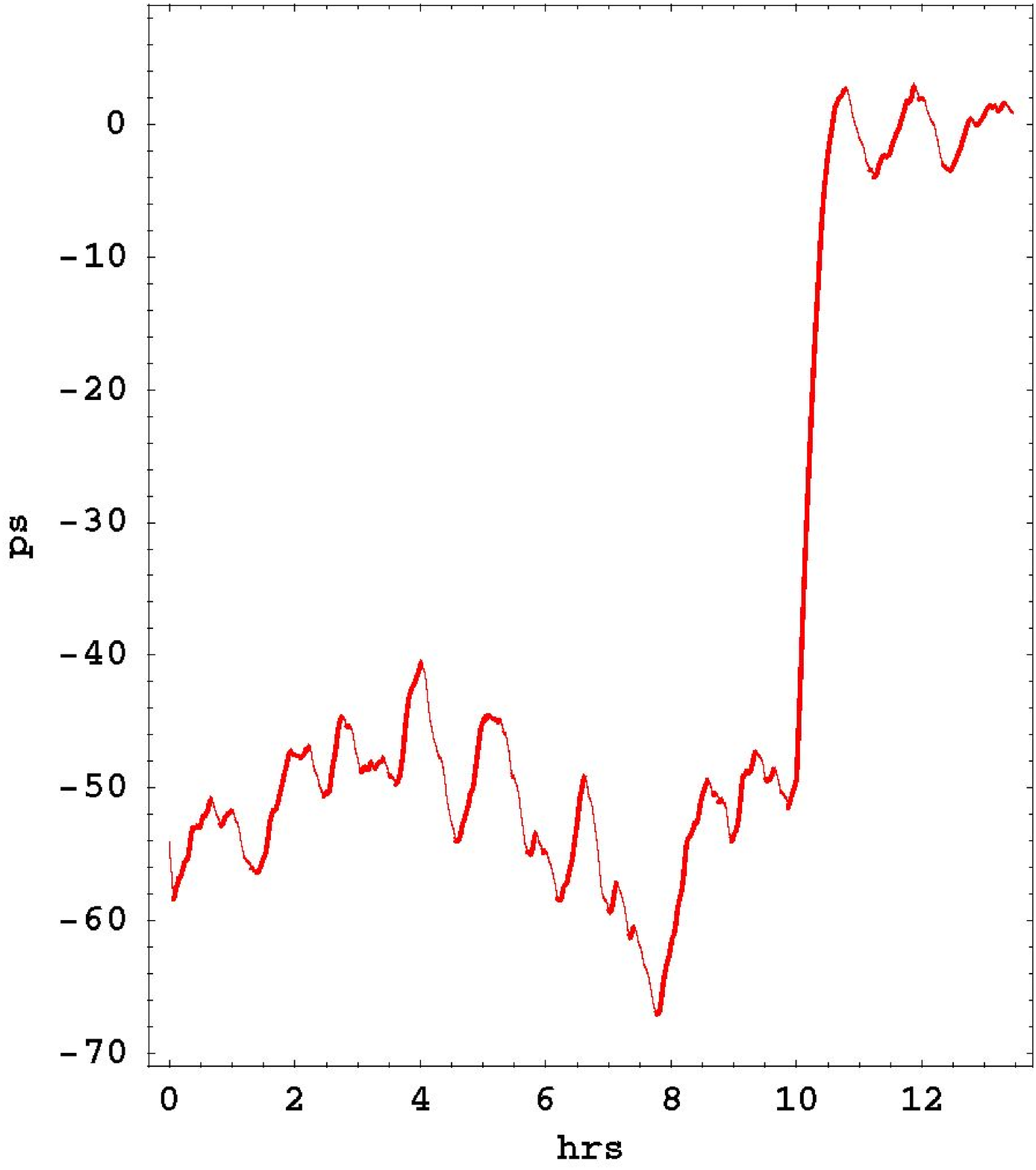}

\vspace*{-14pt}
\caption{\small{The detector arm was formed into a loop at approximately  10:00hrs local time. With the system still operating time averaging causes the trace  to interpolate during this procedure, as shown. This looping  effect is equivalent to having ${\bf v}\,{=}\,{0}$, which defines the value of $\Delta \tau$. In plotting the times here the zero time is set so that then  $\Delta \tau\,{=}\,0$. Now the detector is calibrated, and the times in this figure are absolute times. The times are the N to S travel time subtracted from the shorter S to N travel time, and hence are negative numbers.  This demonstrates that the flow of space past the Earth is essentially from south to north, as shown in Fig.~\ref{fig:Detector}. When the arms are straight, as before 10:00hrs we see that on average the two travel times differ by some 55\,ps.  This looping effect is a critical test for the detector.  It clearly shows the effect of absolute motion upon the RF travel times. As well we see Earth rotation, wave and converter noise effects before 10:00hrs, and converter noise and some small signal after 10:00hrs, caused by an imperfect circle. From this data (\ref{eqnum:Dec4}) and (\ref{eqnum:Dec5}) give  
$\delta\,{=}\,\mbox{72}^\circ$\,S and $v \,{=}\,\mbox{418}$\,km/s.}}
\label{fig:Loop}

\vspace*{-3pt}
\end{figure}

\smallskip\noindent {\bf Fibre Optic Transceivers:}
Fiber-Span AC231-EB-1-3 RF/ Fiber Optic Transceiver (O/E and E/O).  Is a linear ex\-tend\-ed band (5--2000\,MHz) low noise RF fibre optic transceiver for single mode  1.3\,$\mu$m fibre optic wireless systems, with in\-de\-pendent receiver and transmitter. RF interface is a 50$\Omega$  con\-nector and the optical connector is a low reflection FC/APC connector.  Temperature dependence of phase delay is not mea\-sured yet.  The experiment is  operated  in a uniform tem\-perature room, so that phase delays between the two trans\-ceivers cancel to some extent.

\smallskip\noindent {\bf Coaxial Cable:}
Andrews  FSJ1-50A Phase Stabilised 50$\Omega$ Coaxial  Cable. Travel time temperature dependence is 0.026\,ps/m/$^\circ$C. The speed of RF waves in this cable is $c/n\,{=}$ ${=}\,\mbox{0.84}\,c$, arising from the dielectric having refractive index $n\,{=}\,\mbox{1.19}$. As well temperature effects cancel because the two coaxial cables are tied together, and so only temperature differences between adjacent regions of the cables can have any effect.  If such temperature differences are $<$1$^\circ$C, then temperature generated timing errors from this source should be $<$0.3\,ps for the  10\,m.

\smallskip\noindent {\bf Optical Fibre:}
Sumitomo Electric Industries Ind. Ltd Japan  Phase Stabilised Optical Fibre (PSOF) --- single mode.
Uses  
Liquid Crystal Polymer (LCP) coated single mode optical fibre, with this coating designed to make the travel time tem\-perature dependence $<$0.002\,ps/m/$^\circ$C very small compared to normal fibres (0.07\,ps/m/$^\circ$C).  As well temperature effects cancel because the two optical fibres are tied together, and so only temperature differences between adjacent regions of the fibres can have any effect.  If such temperature differences are $<$1$^\circ$C, then temperature generated timing errors from this source should be $<$0.02\,ps for the 10\,m. Now only Furu\-kawa Electric Ind. Ltd Japan manufacturers PSOF.

Photographs of the Flinders detector are shown in Fig.\ref{fig:Expta}.  Because of the new timing technology the detector is now small enough to permit the looping of the detector arm as shown in Fig.~\ref{fig:Exptb}. This enables a key test to be performed as in the loop configuration the signal should disappear, as then the device acts as though it were located at rest in space, because the actual effects of the absolute motion cancel.   The striking results from this test are shown in
 Fig.~\ref{fig:Loop}. As well this key test also provides a means of calibrating the detector.

\subsection{All-optical  detector}\label{subsect:43}

The unique optical fibre effect permits an even more com\-pact gravitational wave detector. This would be an all-optical system 1st order in v/c device, with light passing through vacuum, or just air, as well as optical fibres. The travel time through the fibres is, as above, unaffected by orientation of the device, while the propagation time through the vacuum is affected by orientation, as the device is moving through the local space.

In this system the relative time differences can be mea\-sured using optical interference of the light from the vacuum and fibre components.  Then it is easy to see that the vacuum path length needs only be some 5\,cm.  This makes the con\-struction of a three orthogonal arm even simpler.  It would be a cheap bench-top box.  In which case many of these devices could be put into operation around the Earth, and in space, to observe the new spatial-flow physics, with special emphasis on correlation studies.  These can be used to observe the spatial extent of the fluctuations. As well space-probe based systems could observe special effects in the flow pattern associated with the Earth-Moon system; these effects are caused by the $\alpha$-dependent dynamics in (\ref{eqnum:E1}).

\begin{figure}[t]
\vspace*{-1.2mm}\hspace{-20mm}\includegraphics[scale=0.6]{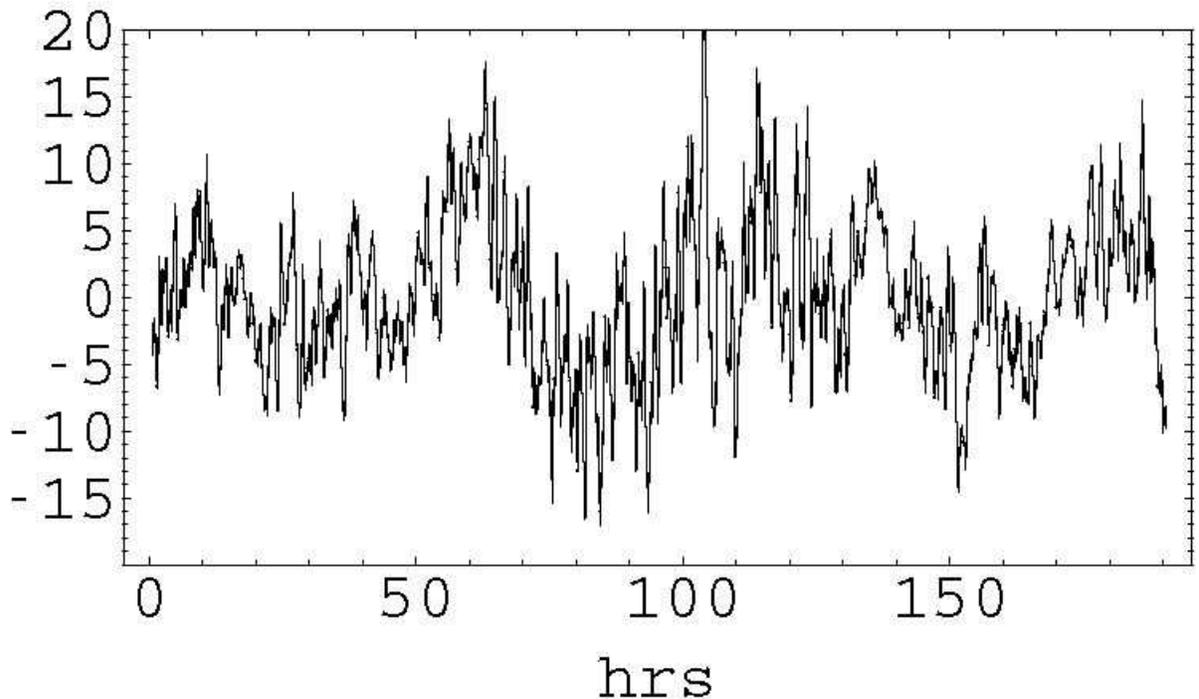}\rule{-10mm}{0pt}

\vspace*{-3.2mm}
\caption{\small   RF travel time variations in picoseconds (ps) for RF waves to travel through, effectively, 10 meters of coaxial cable orientated in a NS direction. 
The data is plotted against local Adelaide time for the days August 18--25, 2006.  The zero of the travel time variations is  arbitrary.   Long term  temperature related drifts over these 7 days have been removed by fitting a low order polynomial to the original data and subtracting the  best fit. The data shows fluctuations identified as earth rotation effect and  gravitational waves.
 These fluctuations exceed those from timing errors in the detector.}  
\label{fig:VaultData}

\vspace*{0mm}
\end{figure}


\begin{figure}[t]
\hspace{40mm}\includegraphics[scale=0.23]{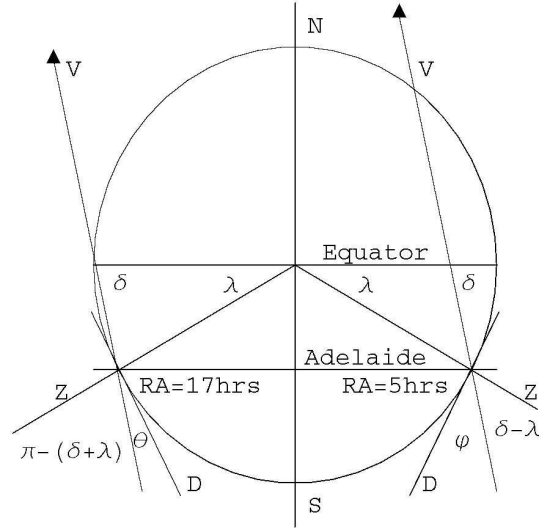}

\vspace*{-3.2mm}
\caption{Profile of Earth, showing NS axis, at Adelaide local sidereal time of RA $\approx $\,5\,hrs (on RHS) and at RA $\approx$\,17\,hrs (on LHS). Adelaide has latitude $\lambda\,{=}\,\mbox{38}^\circ$\,S.   $Z$ is the local zenith, and the detector arm has horizontal local NS direction D. The  flow of space past the Earth has average velocity ${\bf v}$.  The average direction, $-{\bf v}$, of motion of the Earth through local 3-space has RA $ {\approx}\,$5\,hrs  and Declination $\delta \,{\approx}\,\mbox{70}^\circ$S. The angle of inclination of the detector arm D to the  direction $-{\bf v}$   is  
$\phi\,{=}\,\frac{\pi}{2}\,{-}\,\delta\,{+}\,\lambda$ and 
$\theta\,{=}\,\delta\,{+}\,\lambda\,{-}\,\frac{\pi}{2}$  at these two RA, respectively. As the Earth rotates the inclination angle changes from a minimum of $\theta$ to a maximum of $\phi$, which causes the dominant ``dip'' effect in, say, Fig.~\ref{fig:STEffect}.  The gravitational wave effect is the change of  direction and magnitude of the flow velocity ${\bf v}$, which causes the fluctuations in,  say, Fig.~\ref{fig:STEffect}.  
The lati\-tude of Mt.\,Wilson is  34$^\circ$\,N, and  so its latitude almost
mirrors that of Adelaide. This is relevant to the comparison in Fig.~\ref{fig:SeptPlot}.
\label{fig:Detector}}

\vspace*{-2.3mm}
\end{figure}

\vspace*{-1pt}
\subsection{Results from the Flinders detector}\label{subsect:44}

Results from the detector are shown in Fig.~\ref{fig:VaultData}. There the time variations in picoseconds are plotted against local Ade\-l\-aide time. The times have an arbitrary zero offset.   However most significantly we see $\sim$24\,hr variations in the travel time, as also seen by De Witte.  We also see variations in the times and magnitudes from day to day and within each day. These are the wave effects although as well a com\-ponent of these is probably also coming from temperature change effects in the optical fibre transceivers.  In time the inst\-rument will be improved and optimised.  But we are cer\-tainly seeing the evidence of absolute motion, namely the detection of the velocity field, as well as fluctuations in that velocity.  To understand the daily variations we show in Fig.~\ref{fig:Detector} the orientation of the detector arm relative to the Earth rotation axis and the Miller flow direction, at two key local sidereal times.
So we now have a very inexpensive gravitational wave detector sufficiently small that even a coaxial-cable  three-arm detector could easily be located within a building. Three orthogonal arms permit a complete measurement of the spa\-tial flow velocity.  Operating such a device over a year or so will permit the extraction of the Sun in-flow component and the Earth in-flow component, as well as a detailed study of the wave effects. 

\vspace*{-1pt}
\subsection{Right ascension}\label{subsect:45}

\vspace*{-1pt}
 The sidereal effect has been well established, as shown in Fig.~\ref{fig:DeWitteST} for both the De Witte and Flinders data.
 Fig.~\ref{fig:MillerAz} clearly 
shows that effect also for the Miller data.  None of the other an\-isotropy experiments took data for a sufficiently long enough time to demonstrate this effect, although their results are consistent with the Right Ascension and Declination found by the Miller, De Witte and Flinders experiments.  From some 25 days of data in August 2006, the local Adel\-aide time for the largest travel-time difference is approxima\-tely 10$\pm$2\,hrs. This corresponds to a local sidereal time of 17.5$\pm$2\,hrs. According to the Miller convention we give the direction of the velocity vector of the Earth's motion through the space, which then has Right Ascension 5.5$\pm$2\,hrs. This agrees remarkably well with the Miller and De Witte Right  Ascension determinations, as discussed above.  A one hour change in RA corresponds to a 15$^\circ$ change in direction at the equator. However because the declination, to be determin\-ed next, is as large as some 70$^\circ$, the actual RA variation  of $\pm$2\,hrs, corresponds to an angle variation of some $\pm\mbox{10}^\circ$ at that declination.  On occasions there was no discernible unique maximum travel  time difference; this happens when the declination is fluctuating near 90$^\circ$, for then the RA be\-comes ill-defined.
 
\begin{figure}[t]
\hspace{0mm}\includegraphics[scale=0.5]{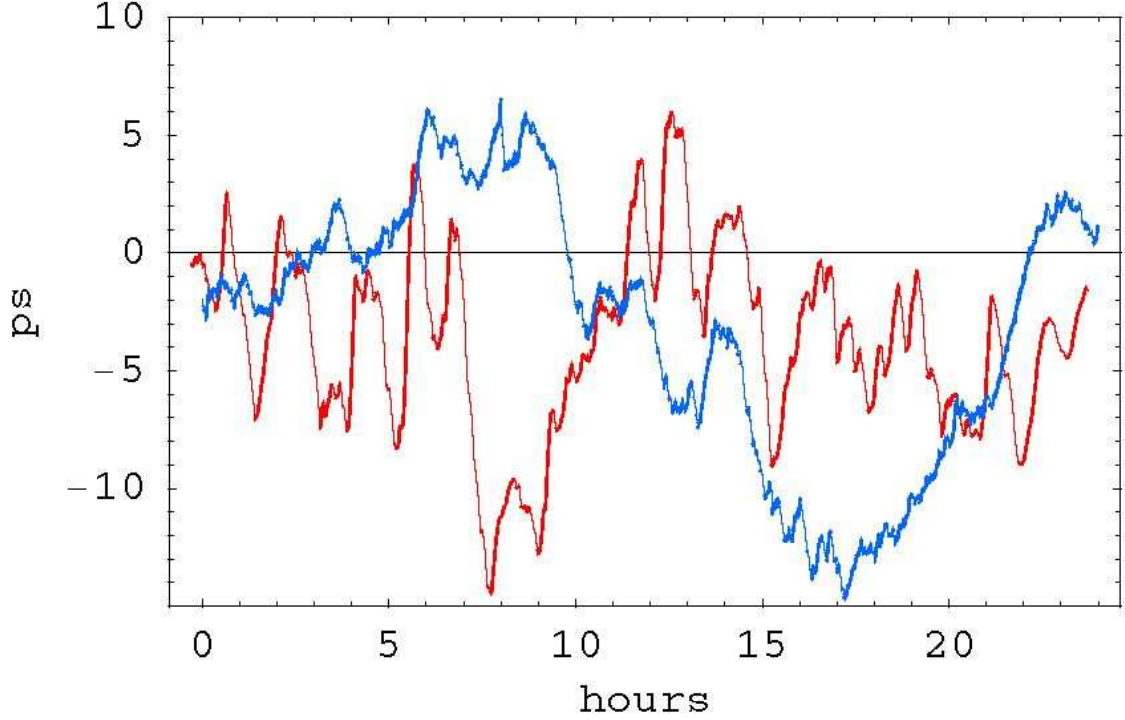}

\vspace*{-1.5mm}
\caption{\small{ The superimposed plots show the sidereal time effect. The plot (blue)  with the minimum at approximately 17\,hrs local Adelaide time is from June 9, 2006, while the plot  (red) with the minimum at ap\-pro\-xi\-ma\-tely 8\,hrs local time is from August 23, 2006. We see that the minimum has moved forward in time by approximately 9\,hrs.  The expected shift for this 65 day difference, assum\-ing no wave effects, is 4.3\,hrs, but     the wave effects shift the RA by some $\pm$2\,hrs on each day as also shown in Fig.~\ref{fig:DeWitteST}. This sidereal time shift is a critical test for the confirmation of the detector.  Miller also detected variations of that magnitude as shown in Fig.~\ref{fig:MillerAz}. The August 23 data is also shown in Fig.~\ref{fig:SeptPlot}, but there plotted against local sidereal time for comparison with the De Witte and Miller data.}}
\label{fig:STEffect}

\vspace*{1.0mm}
\end{figure}

\begin{figure}[p]

\begin{center}
\vspace*{-.2mm}\hspace{0mm}\includegraphics[scale=0.21]{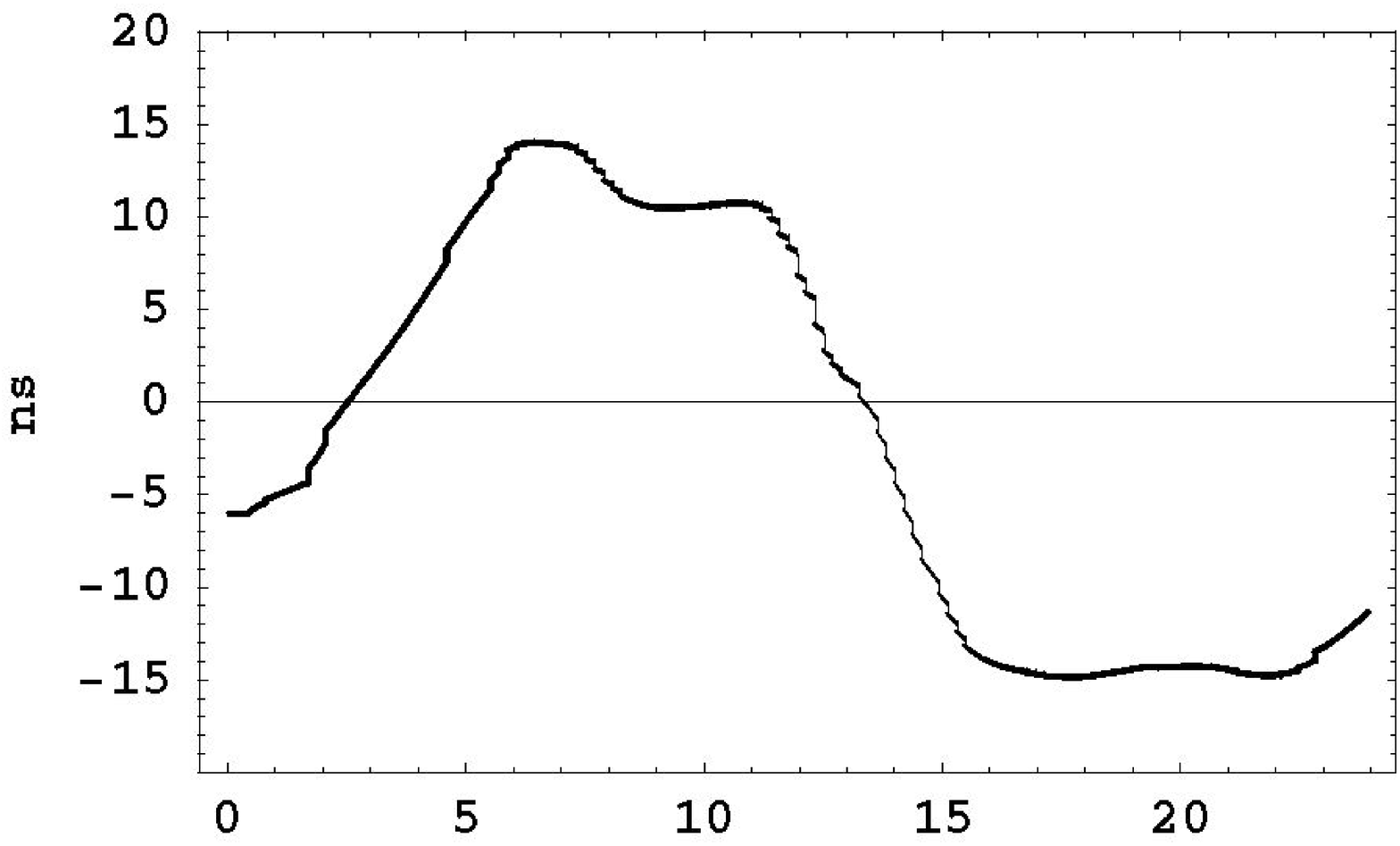}

\vspace*{2.0mm}\hspace{4.4mm}\includegraphics[scale=0.83]{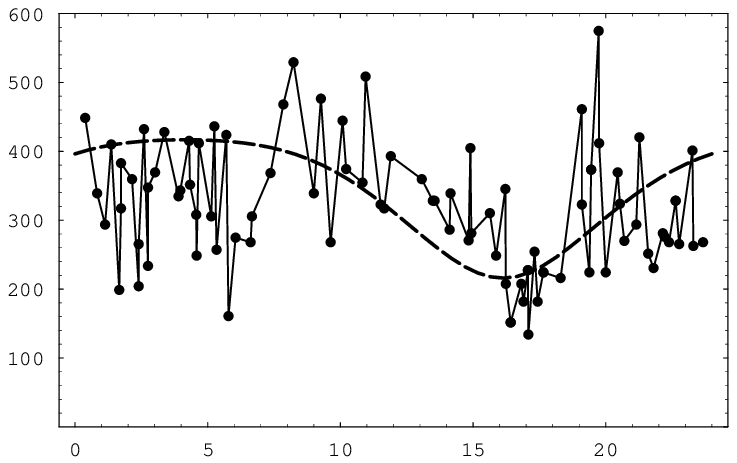}

\vspace*{2.0mm}\hspace{-1.0mm}\includegraphics[scale=0.225]{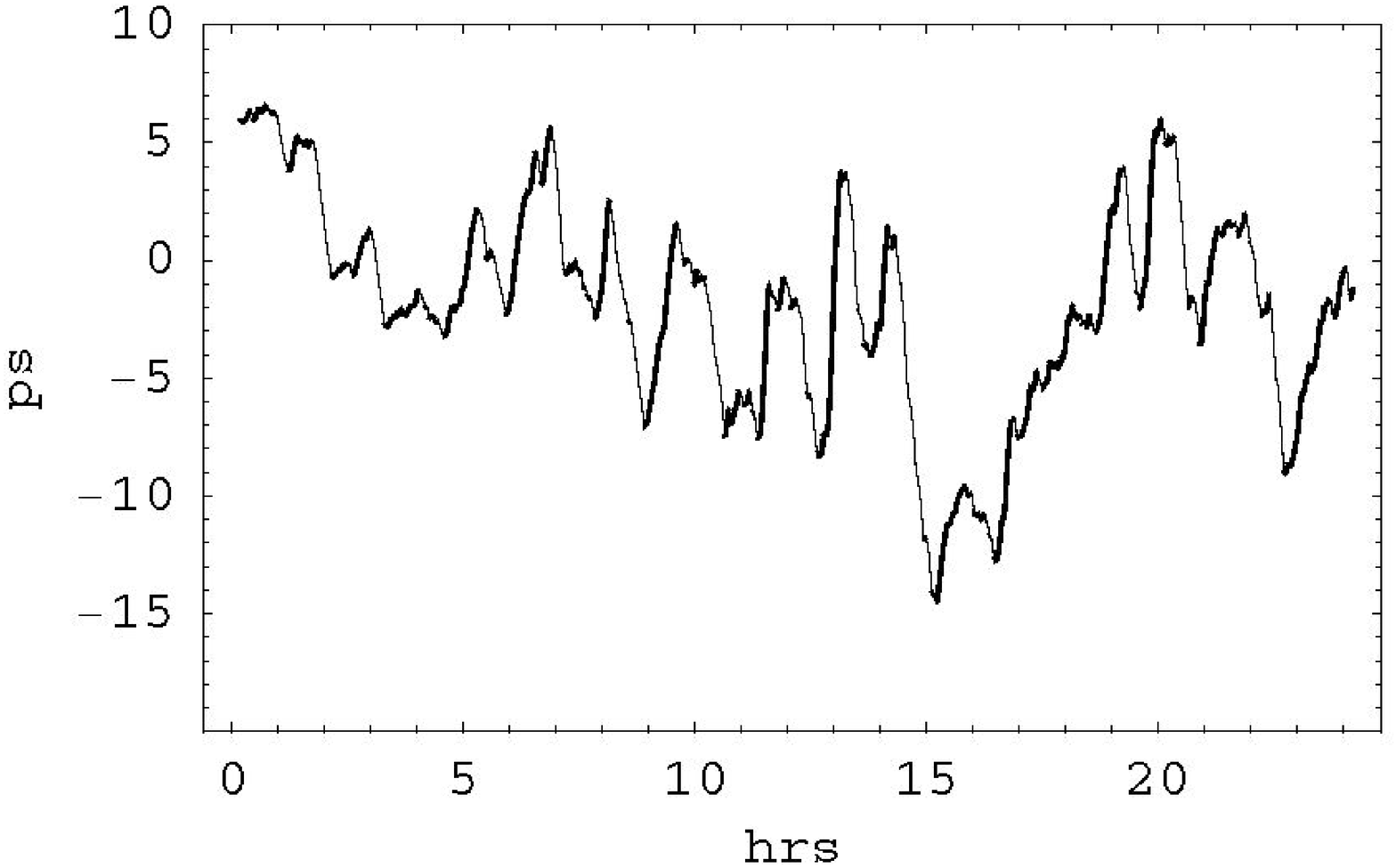}
\end{center}

\vspace*{-4.4mm}
\caption{\small{ {\bf Top:} De Witte data, with sign reversed,  from the first sidereal day in 
Fig.~\ref{fig:DeWittetimes}. This data gives a speed of approximately 430km/s. The  data appears to have been averaged over more than 1hr, but still shows wave effects. 
{\bf Middle:} Absolute projected speeds $v_P$ in the Miller experiment  plotted against sidereal time in hours  for a composite day collected over a number of days in Sep\-tem\-ber 1925.  Speed data like this comes from the fits as in Fig.~\ref{fig:Miller} using the Special Relativity calibration in (\ref{eqnum:e6}). Maximum projected speed is 417\,km/s, as given in  \cite{C1,C10,DeWitte}.
The data shows consider-\protect able fluctuations.  The dashed curve shows the non-fluctuating  variation expected over one day as the Earth rotates, causing the projection onto  the plane of the interferometer of the  velocity of the average direction of the space flow to change.  If the data was plotted against solar time the form is shifted by many hours.  Note that the min/max occur at approximately 5\,hrs and 17\,hrs, as also seen by De Witte and  the new experiment herein. The cor\-res\-pond\-ing variation of the azimuthal phase $\psi$ from Fig.~\ref{fig:Miller}  is shown in 
Fig.~\ref{fig:MillerAz}.   {\bf Bottom:}  Data from the new experiment for one sidereal day on approximately August 23.  We see similar variation with sidereal time, and also similar  wave structure.  This  data has been averaged over a running 1hr time interval to more closely match the time resolution of the Miller experiment. These fluctuations are believed to be real wave phenomena, predicted by the new theory of space \cite{C11}.  The new experiment gives a speed of 418\,km/s. We see remarkable agreement between all three experiments.}}
\label{fig:SeptPlot}\end{figure}

\subsection{Declination and speed}\label{subsect:46}

\vspace*{-1pt}
Because the prototype detector has only one arm, rather than the ideal case of three orthogonal arms, to determine the de\-cli\-na\-tion and speed we assume here that the flow is uni\-form and time-independent, and use the changing difference in travel times between the two main coaxial cables.  Consider  Fig.~\ref{fig:FUGWD} showing  the detector schematic layout and  Fig.~\ref{fig:Detector}  showing  \,the \,various \,angles. \,The \,travel \,time \,in \,one \,of \,the

\noindent
circuits  is given by
\vspace*{-2pt}
\begin{equation}
t_1=\tau_1+\frac{L_1}{v_c-v\cos(\Phi)}
\label{eqnum:Dec1}\end{equation}

\vspace*{-2pt}\noindent
and that in the other arm by
\vspace*{-2pt}
\begin{equation}
t_2=\tau_1+\frac{L_1}{v_c+v\cos(\Phi)}
\label{eqnum:Dec2}\end{equation}

\vspace*{-2pt}\noindent
where $\Phi$ is the angle between the detector direction and the flow velocity ${\bf v}$,
$v_c$ is the speed of radio frequency (RF) el\-ec\-t\-ro\-magnetic waves in the coaxial cable when ${\bf v}\,{=}\,{0}$, namely $v_c\,{=}\,c/n$ where $n$ is the refractive index of the dielectric in the coaxial cable,
and $v$ is the change in that speed caused by the abso\-lu\-te motion of the coaxial cables through space, when the cable is parallel to ${\bf v}$. The factor of $\cos(\Phi)$ is just the projection of   ${\bf v}$ onto the cable direction.  The difference in signs in  (\ref{eqnum:Dec1}) and  (\ref{eqnum:Dec2}) arises from the RF waves  travelling in opposite directions in the two main coaxial cables.  The distance $L_1$ is the arm length of the coaxial cable from A to B, and  $L_2$ is that  from C to D.   The constant times $\tau_1$ and $\tau_2$ are travel times arising from the optical fibres,  the converters, and the coaxial cable lengths not included in $L_1$ and $L_2$,  particularly the  optical fibre travel times. which is the key to the new detector.  The effect of the two  shorter coaxial cable  sections  in each arm are included in $\tau_1$ and $\tau_2$ because the absolute motion effects from these arms is additive, as the RF travels in opposite directions through them, and so only contributes at 2nd order.   

Now the experiment involves first the measurement of the difference $\Delta t=t_1-t_2$, giving 
\begin{equation}
\!\,\begin{array}{rr}
\displaystyle
\Delta t=\tau_1- \tau_2+\frac{L_1}{v_c {-} v\cos(\Phi)}-
\frac{L_2}{v_c{+} v\cos(\Phi)} \,\approx\\[+10pt]
\displaystyle
\approx\,\Delta \tau+ (L_1+L_2)\cos(\Phi)\,\frac{v}{v_c^2}+\dots
\label{eqnum:Dec3}
\end{array}\!\!
\end{equation}

\vspace*{-2pt}\noindent
on expanding to  lowest order in $v/v_c$, and where 
$\Delta \tau \,{\equiv}$
${\equiv}\, \tau_1\,{-}\,\tau_2\,{+}\,
{\frac{\displaystyle L_1 {-} L_2}{\displaystyle v_c}}$. Eqn.~(\ref{eqnum:Dec3}) is the key to the operation of the detector. We see that the effective arm length is $L\,{=}$ ${=}\,L_1{+} L_2\,{=}\,\mbox{10}$\,m. Over time  the velocity vector ${\bf v}$ changes, caused by the wave effects and also by the Earth's orbital velocity about the Sun changing direction, and as well  the Earth rotates on its axis. Both of these effects cause $v$ and the angle $\Phi$ to change.  However over a period of  a day and ignoring wave effects we can assume that $v$ is unchanging. Then we can determine a declination $\delta$ and the speed $v$ by (i) measuring the maximum and minimum values of $\Delta t$ over a day, which occur  approximately 12 hours apart, and (ii)~de\-termine $\Delta \tau$, which is the time difference when $v\,{=}\,0$, and this is easily measured by putting the detector arm into a circular loop, as shown in Fig.~\ref{fig:Exptb}, so that  absolute motion effects cancel, at least to 1st order in $v/v_c$.  Now from 
Fig.~\ref{fig:Detector}  we \rule{-.3pt}{0pt}see \rule{-.3pt}{0pt}that \rule{-.3pt}{0pt}the \rule{-.3pt}{0pt}maximum \rule{-.3pt}{0pt}travel \rule{-.3pt}{0pt}time \rule{-.3pt}{0pt}difference $\!\Delta t_{max}\!$ occurs when $\Phi\,{=}\, \theta\,{=}\,\lambda\,{+}\,\delta\,{-}\,\frac{\pi}{2}$ in (\ref{eqnum:Dec3}), and the min\-i\-mum $\Delta t_{min}$ when   $\Phi{=}\phi{=} \lambda{-}\delta{+}\frac{\pi}{2}$,  12 hours later.  Then the declination  $\delta $ may \rule{-.3pt}{0pt}be \rule{-.3pt}{0pt}determined \rule{-.3pt}{0pt}by \rule{-.3pt}{0pt}numerically \rule{-.3pt}{0pt}solving \rule{-.3pt}{0pt}the \rule{-.3pt}{0pt}transcenden\-tal equation which follows from these two times from (\ref{eqnum:Dec3})
\vspace*{-2pt}
\begin{equation}
\frac{\cos(\lambda+\delta-\frac{\pi}{2})}{ \cos(\lambda-\delta+\frac{\pi}{2})}=\frac{\Delta t_{max}-\Delta \tau}{\Delta t_{min}-\Delta \tau}\,.
\label{eqnum:Dec4}\end{equation}

\vspace*{-2pt}
Subsequently the speed $v$ is obtained from 
\vspace*{-2pt}
\begin{equation}
v=\frac{(\Delta t_{max}-\Delta t_{min})\,v_c^2}{L\,
\bigl(\cos(\lambda+\delta-\frac{\pi}{2})- 
\cos(\lambda-\delta+\frac{\pi}{2})\bigr)\rule{0pt}{9.5pt}}\,.
\label{eqnum:Dec5}\end{equation} 

\vspace*{-2pt}
In Fig.~\ref{fig:Loop} we show the travel time variations for  Sep\-tem\-ber 19, 2006. The detector arm was formed into a loop at approximately  10:00\,hrs local time, with the system still op\-er\-ating: time averaging causes the trace  to interpolate during this procedure, as shown. This looping  effect is equivalent to having ${\bf v}\,{=}\,{0}$, which defines the value of $\Delta \tau$. In plotting the times in Fig.~\ref{fig:Loop} the zero time is set so that then  $\Delta \tau\,{=}\,0$. When the arms are straight, as before 10:00\,hrs we see that on average the travel times are some 55\,ps different: this is because the RF wave travelling S to N is now faster than the RF wave travelling from N to S.  The times are negative because the longer S to N time is subtracted from the shorter N to S travel time in the DSO. As well we see the daily variation as the Earth rotates, showing in particular the maximum effect at approximately 8:00\,hrs local time (approximately 15hrs sidereal time) as shown for the three experiments in Fig.~\ref{fig:SeptPlot}, as well as wave and converter noise. The trace after 10:00\,hrs should be flat --- but the variations seen are coming from noise effects in the converters as well as some small signal arising from the loop not being formed into a perfect circle.  Taking $\Delta t_{max}\,{=}{-}\mbox{63}$\,ps  and $\Delta t_{min}\,{=}{-}\mbox{40}$\,ps  from Fig.~\ref{fig:Loop}, (\ref{eqnum:Dec4}) and (\ref{eqnum:Dec5}) give  $\delta\,{=}\,\mbox{72}^\circ$\,S and $v \,{=}\,\mbox{418}$\,km/s.  This is in extraordinary agreement with the Miller results for September 1925.

We can also  analyse the De Witte data.  We have $L\,{=}$ ${=}\,\mbox{3.0}$\,km, $v_c\,{=}\,\mbox{200,000}$\,km/s, from Fig.~\ref{fig:DeWittetimes} \newline $\Delta t_{max}{-}\Delta t_{min}\,{\approx}$ ${\approx}\, \mbox{25}$\,ns, and the latitude of Brussels is $\lambda\,{=}\,\mbox{51}^\circ$\,N. There is not sufficient De Witte data to determine the declination of ${\bf v}$ on the days when the data was taken.  Miller found that the declination varied from approximately 60$^\circ$\,S to 80$^\circ$\,S, depending on the month. The dates for the De Witte data in Fig.~\ref{fig:DeWittetimes} are not known but, for example, a declination of $\delta\,{=}\,\mbox{60}^\circ$ gives $v\,{=}\,\mbox{430}$\,km/s.

\vspace*{-1pt}
\subsection{Gravity and gravitational waves}\label{sect:gravity}

\vspace*{-1pt}
We have seen that as well as the effect of the Earth rotation relative to the stars, as previously shown by the data from Michelson-Morley, Illingworth,  Joos, Jaseja {\it el al.}, Torr and Kolen, Mil\-ler,  and De Witte and the data from the new experiment herein, \rule{-.2pt}{0pt}there \rule{-.2pt}{0pt}is \rule{-.2pt}{0pt}also \rule{-.2pt}{0pt}from \rule{-.2pt}{0pt}the \rule{-.2pt}{0pt}experimental \rule{-.2pt}{0pt}data \rule{-.2pt}{0pt}of \rule{-.2pt}{0pt}Michelson-Morley, Miller, Torr and Kolen,   De Witte and  from the new experiment, evidence of turbulence in  this  flow of space past  the Earth.   This all points to the flow velocity field ${\bf v}({\bf r},t)$ having a time dependence over an above that caused simply because observations are taken from the rotating Earth.  As we shall now show this turbu\-lence is what is conventionally called ``gravitational waves'', as already noted \cite{C11,C9,C10}.  To do this we briefly review the new dynamical theory  of 3-space, following \cite{BHoles}, although it has been extensively dis\-cussed in the related literature. In the limit of zero vorticity for ${\bf v}({\bf r},t)$ its dynamics is deter\-mined~by
\vspace*{-4pt}
\begin{equation}
\begin{array}{ll}
\displaystyle
\nabla\cdot\left(\frac{\partial {\bf v} }{\partial t}+
({\bf v}\cdot{\bf \nabla}){\bf v}\right)+\\[+10pt]
\displaystyle
\qquad\qquad+\, \frac{\alpha}{8}\left((\mathrm{tr} D)^2-
\mathrm{tr}(D^2)\right)= 
-4\pi G\rho\,,
\end{array}
\label{eqnum:E1}
\end{equation}

\vspace*{-2pt}\noindent
where $\rho$ is the effective matter/energy density, and where 
\vspace*{-2pt}
\begin{equation} D_{ij}=\frac{1}{2}\left(\frac{\partial v_i}{\partial x_j}+
\frac{\partial v_j}{\partial x_i}\right).
\label{eqnum:E2}\end{equation}

\vspace*{-2pt}
Most significantly data from the bore hole $g$ anomaly  and from the systematics of galactic supermassive black hole shows that $\alpha\,{\approx}\,\mbox{1/137}$ is the fine structure constant known from quantum theory 
[21--24].  Now the Dirac equa\-tion  uniq\-ue\-ly  couples to this dynamical 3-space, according to \cite{BHoles} 
\vspace*{-2pt}
\begin{equation}
i\hbar\frac{\partial \psi}{\partial t}\,{=}{-}i\hbar
\biggl(  c\vec{ \alpha}{\cdot}\nabla {+}\, {\bf
v}{\cdot}\nabla{+}\frac{1}{2}\,
\nabla{\cdot}{\bf v} 
\biggr)\psi{+}\,\beta m c^2\psi
\label{eqnum:12}\end{equation}

\vspace*{-2pt}\noindent
where $\vec{\alpha}$ and $\beta$ are the usual Dirac matrices. We can compute the acceleration of a localised spinor wave packet  accord-
ing~to
\vspace*{-7pt}
\begin{equation}
{\bf g}\equiv\frac{d^{ 2}}{dt^2}\bigl(\psi(t),\,{\bf r}\,\psi(t)\bigr)
\label{eqnum:E11}\end{equation}
With ${\bf v}_R\,{=}\,{\bf v}_0\,{-}\,{\bf v}$   the velocity of the wave packet rela\-ti\-ve to the local space, as ${\bf v}_0$ is  the velocity relative to the em\-bedd\-ing space\footnote{See \cite{BHoles} for a detailed explanation of  the embedding space concept.}$\!$, and we obtain
\vspace*{-2pt}
\begin{equation}
{\bf g}\,{=}\,\displaystyle{\frac{\partial {\bf v}}{\partial t}}{+}({\bf v}{\cdot}{\bf \nabla}){\bf
v}{+}({\bf \nabla}{\times}{\bf v}){\times}
{\bf v}_{\!R}^{\phantom{0}}{-}\frac{{\bf
v}_{\!R}}{1{-}{\frac{{\bf v}_R^2}{c^2\rule{0pt}{5.5pt}}}}
\frac{1}{2}\frac{d}{dt}\!\biggl(\!\frac{{\bf v}_R^2}{c^2}\!\biggr)\!\label{eqnum:E13a}
\end{equation}

\vspace*{-2pt}\noindent
which gives the acceleration of quantum matter caused by the inhomogeneities and time-dependencies of ${\bf v}({\bf r},t)$. It has  a term which limits the speed of the wave packet relative to space to be $<\!c$.  Hence we see that the phenomenon of gravity, including the Equivalence Principle,  has been derived from a deeper theory.  Apart from the vorticity\footnote{The vorticity term explains the Lense-Thirring effect \cite{GPB}.} and relativistic terms in (\ref{eqnum:E13a}) the quantum matter acceleration is the same as that of the structured 3-space \cite{BHoles,Schrod}.  

We can now show how this leads to both the spacetime mathematical construct and that the geodesic for matter worldlines in that spacetime is equivalent  to trajectories from  (\ref{eqnum:E13a}).  First we note that (\ref{eqnum:E13a}) may be obtained by extremising the time-dilated elapsed time 
\vspace*{-5pt}
\begin{equation}
\tau[{\bf r}_0]=\int dt \left(1-\frac{{\bf v}_R^2}{c^2}\right)^{\!\!1/2}
\label{eqnum:E13}\end{equation}  
with respect to the particle trajectory ${\bf r}_0(t)$ \cite{C11}. This happens because of the Fermat least-time effect for waves: only along the minimal time trajectory do the quantum waves  remain in phase under small variations of the path. This again empha\-si\-ses  that gravity is a quantum wave effect.   We now introduce a spacetime mathematical construct according to the metric
\vspace*{-2pt}
\begin{equation}
ds^2=dt^2 -\frac{\bigl(d{\bf r}-{\bf v}({\bf r},t)\,dt\bigr)^2}{c^2}=
g_{\mu\nu}dx^{\mu}dx^\nu .
\label{eqnum:E14}\end{equation}

\vspace*{-2pt}
Then according to this metric the elapsed time in (\ref{eqnum:E13}) is
\begin{equation}
\tau=\int dt\,\sqrt{g_{\mu\nu}\frac{dx^{\mu}}{dt}\frac{dx^{\nu}}{dt}}\,,
\label{eqnum:E14b}\end{equation}
and the minimisation of  (\ref{eqnum:E14b}) leads to the geodesics of the spacetime, which are thus equivalent to the trajectories from (\ref{eqnum:E13}), namely (\ref{eqnum:E13a}).
Hence by coupling the Dirac spinor dyn\-am\-ics to the space dynamics we derive the geodesic for\-mal\-ism of General Relativity as a quantum effect, but without ref\-er\-ence to the Hilbert-Einstein equations for the induced met\-ric.  \rule{-.4pt}{0pt}Indeed \rule{-.4pt}{0pt}in \rule{-.4pt}{0pt}general \rule{-.4pt}{0pt}the \rule{-.4pt}{0pt}metric \rule{-.4pt}{0pt}of  \rule{-.4pt}{0pt}this \rule{-.4pt}{0pt}induced \rule{-.4pt}{0pt}spacetime will not satisfy  these equations as the dynamical space in\-vol\-ves the $\alpha$-dependent  dynamics, and $\alpha$ is missing from GR\footnote{Why the Schwarzschild metric, nevertheless, works is explained in~\cite{BHoles}.}$\!$.  

Hence so far we have reviewed the new theory of gravity as it emerges within the new physics\footnote{Elsewhere it has been shown that this theory of gravity explains the bore hole anomaly, supermassive black hole systematics, the ``dark matter''  spiral galaxy rotation anomaly effect,  as well as the putative successes of GR, including light bending and gravitational lensing. }$\!$. In explaining gravity we discover that the Newtonian theory is actually flawed: this happened because the motion of planets in the solar system is too special to have permitted Newtonian to model all aspects of the phenomenon of gravity, including that the fundamental dynamical variable is a velocity field and not an acceleration field. 

We now discuss the phenomenon of the so-called ``grav\-i\-ta\-tional waves''.  It may be shown that the metric in (\ref{eqnum:E14}) sat\-isfies the Hilbert-Einstein GR equations, in ``empty'' space, but {\it only} when $\alpha \rightarrow 0$:
\vspace*{-2pt}
\begin{equation}
G_{\mu\nu}\equiv R_{\mu\nu}-\frac{1}{2}\,R g_{\mu\nu}=0\,,
\label{eqnum:GR32}\end{equation}

\vspace*{-2pt}\noindent
where  $G_{\mu\nu}$ is  the Einstein tensor, $R_{\mu\nu}\,{=}\,R^\alpha_{\mu\alpha\nu}$
and $R\,{=}$ ${=}\,g^{\mu\nu}R_{\mu\nu}$ and
$g^{\mu\nu}$ is the matrix inverse of $g_{\mu\nu}$, and  the curvature tensor
is
\begin{equation}
R^\rho_{\mu\sigma\nu}=\Gamma^\rho_{\mu\nu,\sigma}-\Gamma^\rho_{\mu\sigma,\nu}+
\Gamma^\rho_{\alpha\sigma}\Gamma^\alpha_{\mu\nu}-\Gamma^\rho_{\alpha\nu}\Gamma^\alpha_{\mu\sigma},
\label{eqnum:curvature}\end{equation}
where $\Gamma^\alpha_{\mu\sigma}$ is the affine connection
\vspace*{-2pt}
\begin{equation}
\Gamma^\alpha_{\mu\sigma}=\frac{1}{2}\, 
g^{\alpha\nu}\!\left(\frac{\partial g_{\nu\mu}}{\partial x^\sigma}+
\frac{\partial g_{\nu\sigma}}{\partial x^\mu}-
\frac{\partial g_{\mu\sigma}}{\partial x^\nu} \right).
\label{eqnum:affine}\end{equation}

Hence the GR formalism fails on two grounds: (i) it does not include the spatial self-interaction dynamics which has coupling constant  $\alpha$, and (ii) it very effectively obscures the dynamics, for the GR formalism has spuriously introduced the speed of light when it is completely absent from (\ref{eqnum:E1}), except on the RHS when the matter has speed near that of $c$ relative to the space\footnote{See \cite{C11} for a possible generalisation to include vorticity effects and matter related relativistic effects.}$\!$. Now when  wave effects are supposedly extracted from (\ref{eqnum:GR32}), by perturbatively expanding about a background metric, the standard derivation supposedly leads to waves with speed $c$. This derivation must be manifestly incorrect, as the underlying equation   (\ref{eqnum:E1}), even in the limit $\alpha \rightarrow 0$, does not even contain $c$. In fact an analysis of  (\ref{eqnum:E1}) shows that the perturbative wave effects are fluctuations of  ${\bf v}({\bf r},t)$, and travel at approximately that speed, which in the case of the data reported here is some 400\,km/s in the case of earth based detections, i.\,e. 0.1\% of $c$.  
These waves also generate gravitational effects, but only because of the $\alpha$-dependent dynamical effects: when $\alpha\,{\rightarrow}\, 0$ we still have wave effects in the velocity field, but that they produce no  gravitational acceleration effects upon quantum matter.  Of course even in the case of  $\alpha\,{\rightarrow}\, 0$ the velocity field wave effects are detectable by their effects upon EM radiation, as shown by various gas-mode Michelson interferometer and coaxial cable experiments. Amazingly there is evidence that Michelson-Morley \rule{-.3pt}{0pt}actually \rule{-.3pt}{0pt}detected \rule{-.3pt}{0pt}such \rule{-.3pt}{0pt}gravitational \rule{-.3pt}{0pt}waves  as well as the absolute motion effect in 1887, because fluc\-tua\-tions from  day to day of their data shows effects similar to those reported by Miller, Torr and Kolen, De Witte, and the new experiment herein.    Of course if the Michelson interfe\-ro\-meter is operated in vacuum mode  it is totally insensitive to  absolute motion effects and to the accompanying wave effects, as is the case.  This implies that experiments such as the long baseline terrestrial  Michelson interferometers are seriously technically flawed as gravitational wave detectors.  However as well as the various successful experimental tech\-niques  discussed herein for detecting absolute motion and gravitational wave effects a novel technique is that these effects will manifest in the gyroscope precessions observed by the Gravity Probe B satellite experiment  \cite{GPB,GPBwaves}.  

Eqn.~(\ref{eqnum:E1}) determines the dynamical time evolution of the velocity field. However that aspect is more apparent if we write that equation in the integro-differential form
  \begin{equation}
\begin{array}{ll}
\displaystyle
 \frac{\partial {\bf v}}{\partial t}=-\nabla\left(\frac{{\bf v}^2}{2}\right)+\\[+10pt]
\displaystyle
\qquad +\;G\!\int d^{ 3} r^\prime\,
 \frac{\rho_{DM}^{\phantom{0}}({\bf r}^\prime, t)+\rho\,({\bf r}^\prime, t)}{|{\bf r}-{\bf r^\prime}|^3}({\bf r}-{\bf r^\prime})
\end{array}\label{eqnum:E8}
 \end{equation}
in which $\rho_{DM}$ is velocity dependent,
\begin{equation}
\rho_{DM}({\bf r},t)\equiv\frac{\alpha}{32\pi G}\bigl( (\mathrm{tr} D)^2
-\mathrm{tr} (D^2)\bigr)
\,,  
\label{eqn:E7b}\end{equation} 
and is the effective ``dark matter'' density. This shows se\-ver\-al key aspects: (i) there is a local cause for the time de\-pen\-dence from the $\nabla$ term,  and (ii)  a non-local action-at-a-distance effect from the  $\rho_{DM}$  and $\rho$ terms. This is caused by space being essentially a quantum system, so this is better understood as a quantum non-local effect. However (\ref{eqnum:E8}) raises the question of where the observed wave effects come from?  Are they local effects or are they manifestations of distant phenomena? In the latter case we have a new astron\-o\-mical window on the universe.

\markright{R.\,T.\,Cahill. A New  Light-Speed Anisotropy Experiment: Absolute
Motion and Gravitational Waves Detected}

\section{Conclusions}\label{conclusions}

\markright{R.\,T.\,Cahill. A New  Light-Speed Anisotropy Experiment: Absolute
Motion and Gravitational Waves Detected}

\vspace*{-3pt}
 We now have eight experiments that independently and con\-sis\-tently demonstrated (i) the anisotropy of the speed of light, and \rule{-.3pt}{0pt}where \rule{-.3pt}{0pt}the \rule{-.3pt}{0pt}anisotropy  \rule{-.3pt}{0pt}is  \rule{-.3pt}{0pt}quite \rule{-.3pt}{0pt}large, namely 300,000 $\pm$ 400\,km/s, depending on the direction of measurement rela\-ti\-ve to the Milky Way,  (ii) that the direction, given by the Right Ascension and Declination, is now known, being es\-tab\-lished by the Miller, De Witte and Flinders experiments\footnote{Intriguingly this direction is, on average, perpendicular to the plane of the ecliptic. This may be a dynamical consequence of the new theory of space.}$\!$.  The reality of the cosmological meaning of the speed was confirmed by detecting the sidereal time shift over  6 months and more,  (iii) that the relativistic Fitzgerald-Lorentz length contraction is a real effect,  for otherwise the results from the gas-mode interferometers would have not agreed with those from the coaxial cable experiments, (iv)  that Newtonian phys\-ics gives the wrong calibration for the Michelson inter\-ferometer,  which of course is  not surprising, (v) that the observed anisotropy means that these eight experiments have  detected the existence of a 3-space, (vi) that the motion of that 3-space past the Earth displays wave effects at the level of  $\pm$20km/s, as con\-firm\-ed by three  experiments, and possibly present even  in the Michelson-Morley data.  

The Miller experiment was one of the most significant experiments of the 20th century.  It meant that a substructure to reality deeper than space\-time had been revealed, that space\-time was merely a mathematical construct and not an as\-pect of reality.   It meant that the Einstein postulate regard\-ing the invariance of the speed of light was incorrect --- in dis\-agreement with experiment, and had been so from the beginning.  This meant that the Special Relativity effects required a different explanation, and indeed Lorentz had supplied that some 100 years ago: in this it is the absolute motion of systems through the dynamical 3-space that causes SR effects, and which is diametrically opposite to the Ein\-stein formalism.  This has required the generalisation of the Maxwell equations, as first proposed by Hertz in 1888 \cite{Hertz}), and of the  Schr\"{o}dinger and Dirac equations \cite{BHoles,Schrod}. This in turn has lead to a derivation of the phenomenon of gravity, namely that it is caused by the refraction of quantum waves by the inhomogeneities and time dependence of the flowing patterns within space.   That same data has also revealed the in-flow component of space past the Earth towards the Sun \cite{C11}, and which also is revealed by the light bending effect observed by light passing close to the Sun's surface \cite{BHoles}.   This theory of gravity has  in turn lead to an explanation of the so-called ``dark matter'' effect in spiral galaxies \cite{DM}, and to the systematics of black hole masses in spherical star systems \cite{BHoles}, and to the explanation of the bore hole $g$ anomaly \cite{alpha,DM,boreholes}.  These effects have permitted the de\-ve\-lop\-ment of the minimal dynamics of the 3-space, leading to the discovery that the parameter that determines the strength of the spatial self-interaction is none other than the fine structure constant, so hinting at a grand unification of space and the quantum theory, along the lines proposed in \cite{C11}, as an {\it information theoretic} theory of reality.

These developments demonstrate the enormous sig\-nif\-i\-cance of the Miller experiment, and the extraordinary degree to which Miller went in testing and refining his interfero\-meter.  The author is proud to be extending the Miller dis\-coveries by studying in detail the wave effects that are so apparent in his extensive data set.  His work demonstrates the enormous importance of doing novel experiments and doing them well, despite the prevailing prejudices.  It was a tragedy and an injustice that Miller was not recognised for his contributions to physics in his own lifetime; but not everyone is as careful and fastidious with detail as he was.       He was ignored by the physics community simply because in his era it was believed, as it is now, that absolute motion was incompatible with special relativistic effects, and so it was accepted, without any evidence,  that his experiments were wrong.  His experiences showed yet again that few in physics actually accept that it is an evidence based science, as Galileo long ago discovered also to his great cost.  For more than 70 years  this experiment has been ignored, until re\-cent\-ly, but even now discussion of this and related experi\-ments attracts hostile reaction from the physics community. 

The developments reported herein have enormous sig\-ni\-fi\-cance for fundamental physics --- essentially the whole paradigm of 20th century physics collapses.  In particular spacetime is now seen to be no more than a mathematical construct, that no such union of space and time was ever man\-dated by experiment.   The putative successes of Special Re\-la\-tivity can be accommodated by the reality of a dynam\-ic\-al 3-space, with time a distinctly different phenomenon.  But motion of quantum and even classical electromagnetic fields through that dynamical space explain the SR effects.  Lorentz symmetry remains valid, but must be understood as applying only when the space and time coordinates are those arrived at by the Einstein measurement protocol, and which amounts to not making corrections for the effects of absolute motion upon rods and clocks on those measurements.  Nevertheless such coordinates may be used so long as we understand that they lead to a confusion of various related effects.  To correct the Einstein measurement protocol readings one needs only to have each observer use an absolute motion meter, such as the new compact all-optical devices, as well as a rod and clock. 
The fundamental discovery is that for some 100 years physics has failed to realise that a dynamical 3-space exists --- it is observable. This contradicts two previous assumptions about space: Newton asserted that it existed, was unchang\-ing, but not observable, whereas Einstein asserted that 3-space did not exist, could not exist, and so clearly must be unobservable.  The minimal dynamics for this 3-space is now known, and it  immediately explains such effects as the ``dark matter'' spiral galaxy rotation anomaly, novel black holes with non-inverse square law gravitational accelerations, which would appear to offer an explanation for the precoc\-ious formation of spiral galaxies, the bore hole anomaly and the systematics of supermassive black holes, and so on. Dramatically various pieces of data show that the self-interaction constant for  space is the fine structure constant.  However unlike SR, GR turns out to be  flawed but only be\-cause it assumed the correctness of Newtonian gravity.  The self-interaction effects for space make that theory invalid even in the non-relativistic regime --- the famous universal inverse square law of Newtonian gravity is of limited valid\-ity.    Uniquely linking the quantum theory of matter with the dynamical space   shows that gravity is a quantum matter wave \rule{-.3pt}{0pt}effect, so \rule{-.3pt}{0pt}we \rule{-.3pt}{0pt}can't \rule{-.3pt}{0pt}understand \rule{-.3pt}{0pt}gravity \rule{-.3pt}{0pt}without \rule{-.3pt}{0pt}the \rule{-.3pt}{0pt}quan\-tum theory.  As well the dynamics of space is intrinsically non-local, which implies a connectivity of reality that far exceeds any previous notions.

This research is supported by an Australian Research Coun\-cil Discovery Grant {\it Development and Study of a New Theory of Gravity}.  
 
Special thanks to Professor Igor Bray~(Mur\-doch Univ.),   Professor Warren Lawrance (Flinders Univ.), Luit Koert De Jonge (CERN), Tim Cope (Fiber-Span),  Peter Gray (Trio),   Bill Drury and  Dr Lance McCarthy (Flinders Univ.), Tom Goodey (UK), Shizu Bito (Japan), Pete Brown (Mountain Man Graphics), Dr Tim Eastman (Washington), Dr Dmitri Rabounski (New Mexico) and Stephen Crothers (Australia).


\centerline{\rule{0pt}{5pt}\rule{72pt}{0.4pt}\rule[0pt]{0pt}{0pt}}


\begin{thebibliography}{99}\small

\bibitem{C11}  Cahill  R.\,T. Process Physics: 
From information theory to
quan\-tum space and matter.  Nova Science, N.Y., 2005.  

\bibitem{C2}   Cahill R.\,T. and Kitto K. 
{Michelson-Morley experiments re\-vi\-sit\-ed}.
{\it Apeiron}, 2003, v.\,{10}(2), 104--117.  

\bibitem{C1}   Michelson A.\,A. and  Morley E.\,W. {\it Philos. Mag.},
S.\,5, 1887, v.\,24, No.\,151, 449--463.

\bibitem{C4}   Miller D.\,C. {\it Rev. Mod. Phys.},  1933, 
v.\,{5}, 203--242.

\bibitem{Salz} Cahill R.\,T. Process Physics and Whitehead:  
the new science of space and time.
Whitehead 2006 Conference, Salzburg, to be pub. in proceedings, 2006.   

\bibitem{LA} Cahill R.\,T. Process Physics: self-referential information
and experiential reality. \textit{To be pub}.   

\bibitem{C3}  M\"{u}ller H. {\it et al.} Modern Michelson-Morley experiment
using cryogenic optical resonators. {\it Phys. Rev. Lett.}, 2003,
v.\,91(2), 020401-1.

\bibitem{Schrod} Cahill R.\,T. Dynamical  fractal  3-space and the generalised
Schr\"{o}dinger equation: Equivalence Principle and  vorticity effects.
{\it Progress in Physics},  2006, v.\,1, 27--34.

\bibitem{DeWitte}  Cahill R.\,T. {The Roland DeWitte 1991 experiment}.
{\it Progress in Physics}, 2006, v.\,{3}, 60--65.
 
\bibitem{C8}  Torr D.\,G. and Kolen P. 
\textit{Precision Measurements and
Fun\-dam\-ent\-al Constants},  ed. by Taylor B.\,N.\ and
Phillips W.\,D.  Nat.\ Bur.\ Stand.\ (U.S.), Spec.\ 
Pub., 1984, v.\,617, 675--679.

\bibitem{C5}   Illingworth K.\,K. {\it  Phys. Rev.}, 1927, v.\,3,  692--696.

\bibitem{C6}   Joos G. {\it  Ann. der Physik}, 1930, Bd.\,7,  385.

\bibitem{C7}   Jaseja T.\,S. {\it et al.} {\it  Phys. Rev.},
v.\,A133, 1964, 1221.

\bibitem{MMC}  Cahill  R.\,T. {The Michelson and Morley 1887 experiment
and the discovery of absolute motion}.   {\it Progress in Physics}, 
2005, v.\,{3}, 25--29.

\bibitem{AIP} Cahill R.\,T. The Michelson and Morley 1887 experiment
and the discovery of 3-space and absolute motion. {\it Australian Physics},
Jan/Feb 2006, v.\,46, 196--202. 

\bibitem{NPA}  Cahill R.\,T.  The detection of absolute motion: 
from 1887 to 2005. \textit{NPA Proceedings}, 2005, 12--16.   

\bibitem{IE}  Cahill R.\,T. The speed of light and the Einstein legacy:
1905-2005. {\it Infinite Energy}, 2005, v.\,10(60), 28--27.

\bibitem{EP} Cahill R.\,T. The Einstein postulates 1905-2005: a critical
review of the evidence. In: {\it Einstein and Poincar\'{e}: the Physical
Vacuum}, Dvoeglazov V.\,V. (ed.),  Apeiron Publ., 2006.

\bibitem{C9}  Cahill  R.\,T. Quantum foam, gravity and gravitational
waves. arXiv: physics/0312082.

\bibitem{C10}  Cahill R.\,T. Absolute motion and gravitational effects.
{\it Apei\-ron}, 2004, v.\,11(1), 53--111.  

\bibitem{alpha}  Cahill R.\,T.   Gravity, ``dark matter'' and the fine
structure constant. {\it Apeiron}, 2005, v.\,12(2), 144--177.  
      
\bibitem{DM}   Cahill R.\,T.   ``Dark matter'' as a quantum foam in-flow
effect. 
In: {\it Trends in Dark Matter Research},  ed. J.~Val~Blain, Nova Science,
N.Y., 2005, 96--140.   
         
\bibitem{boreholes} Cahill  R.\,T. {3-Space in-flow theory of gravity:
Boreholes,
blackholes and the fine structure constant}.  {\it Progress in Physics},
2006, v.\,{2}, 9--16.    
 
\bibitem{galaxies}   Cahill   R.\,T. {Black holes in elliptical and
spiral galaxies and in 
globular clusters}. {\it Progress in Physics}, 2005, v.\,{3}, 51--56.

\bibitem{BHoles}  Cahill R.\,T.  Black holes and quantum theory: 
the fine structure constant connection. {\it Progress in Physics}, 
2006, v.\,4, 44--50.

\bibitem{Hertz}  Hertz H.  On the fundamental equations of electro-magnetics
for bodies in motion. {\it Wiedemann's Ann.}, 1890, v.\,41, 369;  
El\-ec\-t\-ric waves, collection of scientific papers. Dover, N.Y., 1962.
\bibitem{Fitzgerald}  Fitzgerald G.\,F. {\it Science}, 1889, v.\,13,
420.

\bibitem{Lorentz}   Lorentz H.\,A. Electric phenomena in a system
moving with any velocity less than that of light. In {\it The Principle
of Re\-la\-tivity},  Dover, N.Y., 1952.    

\bibitem{Hicks}  Hicks W.\,M. On the Michelson-Morley experiment relating
to the drift of the ether. {\it Phil. Mag.}, 1902, v.\,3, 9--42. 

\bibitem{GPB} Cahill  R.\,T. Novel Gravity Probe B frame-dragging effect.
{\it Progress in Physics}, 2005, v.\,3, 30--33.
 
\bibitem{GPBwaves}  Cahill R.\,T.  Novel Gravity Probe B  gravitational
wave de\-tec\-tion. arXiv: physics/0408097.
  

 http://www.scieng.flinders.edu.au/cpes/people/cahill$\_$r/ processphysics.html
 
 http://www.mountainman.com.au/process\_physics.

    
\end{thebibliography}
\end{document}